\documentclass[10pt]{article}
\pdfoutput=1
\usepackage[utf8]{inputenc}
\usepackage[pdftex]{geometry}
\usepackage[english]{babel}
\usepackage[authoryear]{natbib}
\usepackage{latexsym,amsmath,amsfonts,amssymb,textcomp,calc,rotating,units,booktabs,url}
\usepackage{graphicx}
\newcommand{\isotope}[2]{\ensuremath{{}^{#2}\text{#1}}}

\newcommand{\df}{\textrm{d}}
\newcommand{\drv}[2]{\frac{\df #1}{\df #2}}
\usepackage{hyperref}
\begin{document}
\thispagestyle{empty}
\begin{center}
\Large Thomas Ruedas\textsuperscript{1,2,3}\\[2ex]
Doris Breuer\textsuperscript{1}\\[5ex]
\textbf{Dynamical effects of multiple impacts:\\Large impacts on a Mars-like planet}\\[5ex]
final version\\[5ex]
9 January 2019\\[10ex]
published in:\\
\textit{Phys. Earth Planet. Inter.} 287, pp.~76--92 (2019)\\[15ex]
\normalsize
\textsuperscript{1}Institute of Planetary Research, German Aerospace Center (DLR), Berlin, Germany\\[5ex]
\textsuperscript{2}Museum f\"ur Naturkunde Berlin, Germany\\[5ex]
\textsuperscript{3}Institute of Planetology, Westf\"alische Wilhelms-Universit\"at, M\"unster, Germany
\rule{0pt}{12pt}
\end{center}
\vfill
\footnotesize The version of record is available at \url{http://dx.doi.org/10.1016/j.pepi.2019.01.003}.\\
This author post-print version is shared under the Creative Commons Attribution Non-Commercial No Derivatives License (CC BY-NC-ND 4.0).
\normalsize
\newpage

\title{Dynamical effects of multiple impacts:\\Large impacts on a Mars-like planet}
\author{Thomas Ruedas\thanks{Corresponding author: T. Ruedas, Institute of Planetary Research, German Aerospace Center (DLR), Berlin, Germany (Thomas.Ruedas@dlr.de)}\\{\footnotesize Institute of Planetary Research, German Aerospace Center (DLR), Berlin, Germany}\\{\footnotesize Museum f\"ur Naturkunde Berlin, Germany}\\{\footnotesize Institute of Planetology, Westf\"alische Wilhelms-Universit\"at, M\"unster, Germany}\\[2ex]
Doris Breuer\\{\footnotesize Institute of Planetary Research, German Aerospace Center (DLR), Berlin, Germany}}
\date{}
\maketitle
\textbf{Highlights}
\begin{itemize}
\item Combined effects are not simply linear superpositions of individual impacts.
\item Many details of impact history are not preserved in the long-term evolution.
\item Global signatures of different impact histories converge in the long term.
\item Lithospheric delamination may prolong melt generation, affect core.
\end{itemize}
\begin{abstract}
The earliest stage of the evolution of a fully assembled planet is profoundly affected by a number of basin-forming impacts large enough to change the dynamics of its deeper interior. These impacts are in some cases quite closely spaced and follow one another in short time intervals, so that their effects interact and result in behavior that may differ from a simple sum of the effects of two individual and isolated impacts. We use two-dimensional models of mantle convection in a Mars-like planet and a simple parameterized representation of the principal effects of impacts to study some of the dynamical effects and interactions of multiple large impacts. In models of only two impacts, we confirm that the dynamical effects of the impacts reinforce each other the closer they are in space and time but that the effects do not always correspond to straightforward superpositions of those of single, isolated impacts. In models with multiple (4--8) impacts with variable sizes, distances, and frequencies, the global response of the mantle is as variable as the impact sequences in the short term, but in the long term the different evolutionary paths converge for several indicator variables such as the mean flow velocity, temperature, or heat flow. Nonetheless, beyond a certain impact frequency and energy, lithospheric instabilities triggered by large impacts occur on a global scale, reinvigorate mantle dynamics for long time spans, and entail a late stage of melt production in addition to the initial melting stage that is not observed in one- or two-impact models. After one or several very large impacts, some lithospheric material may founder and sink to the core--mantle boundary, and if enough of it accumulates there, it enhances the heat flux out of the core for several hundred millions of years, with possible effects on dynamo activity.
\end{abstract}
\begin{flushleft}
Mars; mantle convection; impact
\end{flushleft}

\section{Introduction}
In recent years, the interactions between very large meteorite impacts and planetary interiors have received increasing attention, and several authors have studied a variety of aspects of impact-induced mantle dynamics especially on Mars \citep[e.g.,][]{Rees:etal02,Rees:etal04,WAWatt:etal09,JHRobe:etal09,Ar-HaGh11,Gola:etal11,JHRoAr-Ha12,RuBr17c,RuBr18b}, but also on Mercury \citep{JHRoBa12,Pado:etal17}, Venus \citep[e.g.,][]{Gill:etal16}, Earth \citep[e.g.,][]{ONeil:etal17}, and the Moon \citep{Rolf:etal17}. Most of these studies concentrate on the effects of a single large impact in order to isolate the physical effects more clearly. However, in the earliest history of the terrestrial planets, basin-forming impacts have followed each other in close succession, and at times also in close proximity, which raises the question whether and how the effects of subsequent large impacts on the deep interior of the target planet may influence each other. \citet{JHRobe:etal09,JHRoAr-Ha12} and \citet{Rolf:etal17} have already explored this possibility to some extent for multiple impacts on Mars and the Moon, respectively, and observed that a succession of large impacts leaves a cumulative signature on physical variables such as the heat flow.\par
In this study we consider different successions of impacts, both idealized ones and examples derived from observed martian giant impacts. Our principal goal is to elucidate some more details of how impacts in close temporal or spatial proximity may influence each other and whether the cumulative effects of a succession of several impacts are additive or lead to some sort of saturation, e.g., because the affected mantle is so depleted in fusible mineral components that it ceases to produce melt readily.

\section{Method and model set-up}
The evolution of the martian interior including concomitant processes such as melting is represented as a two-dimensional fully dynamical model using the convection code STAGYY \citep{Tackley96a,Tackley08} with the additions and modifications described in detail in \citet{RuBr17c,RuBr18b} on a two-dimensional spherical annulus grid \citep{HeTa08} with 512\texttimes 128 points. The convection algorithm solves the equations of mass, momentum, and energy conservation in the anelastic, compressible approximation, making use of a detailed petrological model that includes, among others, melting-related density changes and mineral solid-phase transitions. Following the preferred interior models by  \citet{Kono:etal11} and \citet{Rivo:etal11}, we define our models to consist essentially of two layers, i.e., an upper mantle with an olivine-dominated mineralogy and a lower mantle with an assemblage mostly of a high-pressure olivine polymorph and majorite garnet \citep{BeFe97}; there is no basal (pv+fp) layer. The chemical model assumed for the mantle is the one by \citet{WaDr94}, and for the core, we use a sulfur content derived from \citet{Rivo:etal11}. The viscosity of the mantle is described by an extended Arrhenius-type law for diffusion creep of olivine that includes dependencies on temperature $T$, pressure $p$, and chemical composition, in particular water content \citep{HiKo03,YHZhao:etal09}; its value at surface pressure and the initial potential temperature of the mantle for the bulk water and iron contents of unmolten mantle material is also the scaling viscosity of the model. Internal heating is provided by the radioactive decay of \isotope{K}{40}, \isotope{Th}{232}, \isotope{U}{235}, and \isotope{U}{238}, which are tracked with tracer particles and are redistributed by melting, by which they become concentrated in the melt and in the crust forming from it; the movement and partitioning of water is treated in the same manner. The surface and core--mantle boundary (CMB) are implemented as isothermal free-slip boundaries, whereby the temperature at the CMB is allowed to evolve with time according to a simple energy balance and evolution model of the core; the core model largely follows the approach by \citet{Nimm:etal04} and \citet{JPWiNi04} and is described in some detail in \citet[App.~B]{Rued:etal13a}. The model is initialized on the basis of a radially symmetric adiabatic state prevailing at 4.5\,Ga with a core superheated by 150\,K and a surface at 218\,K; the surface temperature is kept constant in all models over the entire evolution, as it is not thought to have changed strongly enough to have had a feedback effect on the interior. The actual initial state is derived from this configuration by setting the supersolidus region to the solidus, extracting the corresponding amount of melt to form an initial crust of $\sim 50$\,km, and redistributing the depletion of the mantle residue randomly throughout the mantle; the latter lets the mantle appear well-mixed and compositionally homogeneous on large scales. The thickness of this primordial crust is not well constrained, but with the extraction threshold and the mantle solidus fixed, it is mostly dependent on temperature. Subsequent regular melting in the course of the evolution is controlled by the extent to which the temperature exceeds the local solidus; if the amount of newly formed melt passes a certain threshold value and has a supersolidus vertical path to the base of the lithosphere, it is assumed to be extractable, and the excess amount is instantaneously moved to the top to build the crust. The extraction threshold has been estimated to lie between 0.1\% \citep[e.g.,][]{McKenzie85b} and 1--2\% \citep[e.g.,][]{Faul01}, and so we chose a value of 0.7\%. The solidus is the same as in \citet[App.~A]{RuBr17c}. The most important model parameters are summarized in Table~\ref{tab:modpar}. Further technical detail may be found in \citet{RuBr17c,RuBr18b} and their supplementary material.\par
The impact process is not included via coupling or interfacing the convection algorithm with a fully dynamical hydrocode simulation, but is described using a combination of simpler analytical or semi-empirical approximations. As the impacts we consider are modeled after real impact basins on Mars, we start from the known diameter of the final crater, $D_\mathrm{f}$ (cf. Table~\ref{tab:impacts}), and use scaling laws for complex craters to work back to determine the diameter of the impactor:
\begin{equation}
D_\mathrm{imp}=\frac{g^{0.28}}{v_{z,\mathrm{imp}}^{0.56}}\left(\frac{\varrho}{\varrho_\mathrm{imp}}\right)^{0.427} 0.6918D_\mathrm{f}^{1.1346}D_\mathrm{sc}^{0.1475}\label{eq:DimpDf}
\end{equation}
\citep[e.g.,][]{Melosh11,RuBr18b}. Here, $v_{z,\mathrm{imp}}=v_\mathrm{imp}\sin\phi$ is the vertical component of the impactor velocity, and $D_\mathrm{sc}$ is the diameter of the transition between simple and complex craters; in the absence of more specific information, the angle $\phi$ is always set to 45\textdegree, which is the most probable impact angle for an isotropic impactor flux \citep[e.g.,][]{GaWe78}. The density and the absolute value of the impactor velocity are set to those of a homogeneous rocky asteroidal object hitting Mars \citep[cf. Table~\ref{tab:modpar}; ][]{Carry12,Ivanov01}.\par
With regard to the energy input from the impact, we follow the approach outlined by \citet{WAWatt:etal09} in many respects, i.e., we base the estimate of the shock pressure on a one-dimensional impedance-match calculation and derive the waste heat from that value as a function of distance from the impact center using the ``inverse-$r$'' pressure decay formula from \citet{Ruedas17a}. The temperature in the shock-heated volume is forced to a value just above the solidus after the production and extraction of the impact melt, for which an upper limit is imposed. The melt extracted is instantaneously added to the top of the model.\par
\begin{table}
\caption{Main model parameters\label{tab:modpar}}
\begin{tabular}{lcl}\hline
\textit{Mars}\\
Planetary radius, $R$&3389.5\,km&\citet{Arch:etal11}\\
Mass&$6.41712\cdot 10^{23}$\,kg&\citet{Kono:etal11,Jacobson10}\\
Surface gravity, $g$&$3.717$\,m/s\textsuperscript{2}&\\
Surface temperature&218\,K&\citet{Catling15}\\
Mantle thickness, $z_\mathrm{m}$&1659.5\,km&after \citet{Rivo:etal11,Kono:etal11}\\
Initial potential temperature, $T_\mathrm{pot}$&1700\,K&\\
Initial core superheating&150\,K&\\
Surface porosity, $\varphi_\mathrm{surf}$&0.2&\citet{Clifford93}\\
Melt extraction threshold&0.007&\\
Simple/complex crater transition, $D_\mathrm{sc}$&5.6\,km&\citet[Tab.~3]{RoHy12b}\\
Bulk silicate Mars Mg\#&0.75&\citet{WaDr94,SRTaMcLe09}\\
Initial bulk silicate water content&36\,ppm&\citet{WaDr94}\\
Present-day K content&305\,ppm&\citet{WaDr94}\\
Present-day Th content&56\,ppb&\citet{WaDr94}\\
Present-day U content&16\,ppb&\citet{WaDr94}\\
Core S content&16\,wt.\%&\citet{Rivo:etal11}\\[1ex]
\textit{Impactor}\\
Density, $\varrho_\mathrm{imp}$&2720\,kg/m\textsuperscript{3}&\citet{Carry12}\\
Velocity, $v_\mathrm{imp}$&9.6\,km/s&\citet{Ivanov01}\\
\hline
\end{tabular}
\end{table}
In addition to a reference model without any impacts, this study comprises two model subsets with a total of 32 models. The first subset consists of models with two impacts of identical size, of which the first strikes at 4\,Ga (i.e., at $t_0=500$\,Myr model time) and whose spatial and temporal separation is varied systematically in order to study possible effects. The impacts of the two series of this subset each have the magnitude of either an Isidis-forming (I) or of a Utopia-forming (U) event, respectively. The spatial separation $\Delta x$ is measured in multiples of the isobaric core diameter $D_\mathrm{ic}$ and is 1, 2, 5, or 10 in the models of either series; the isobaric core diameter is defined based on the position of the inflection point of the ``inverse-$r$'' pressure decay curve from \citet{Ruedas17a} (cf. Eq.~\ref{eq:Dic}). The temporal separation $\Delta t$ is given in multiples of an estimated ``decay time'' $t_\mathrm{d}$ needed for certain dynamical effects caused by a single impact of a given magnitude to decay to levels close to the pre-impact state. We chose the mean flow velocity $v_\mathrm{rms}$ as the main indicator for the dynamical state and consider the decay to be essentially complete when the difference of $v_\mathrm{rms}$ between the disturbed model and the impact-free reference drops below 5\% of the maximum disturbance at the time of impact; the exact choice is somewhat arbitrary, and we rounded the time to full million years for convenience. In the model series, the second impact follows the first at 0.5, 1, and 2 decay times. $\Delta x$ and $\Delta t$ are thus functions of the size of the impact, and as Isidis and Utopia have quite different magnitudes, so have $D_\mathrm{ic}$ and $t_\mathrm{d}$: for Isidis, we have $D_\mathrm{ic}=135.4$\,km and $t_\mathrm{d}=6$\,Myr, for Utopia, we have $D_\mathrm{ic}=385.6$\,km and $t_\mathrm{d}=12$\,Myr. In addition to these two-impact models, each series also includes an additional model with a single impact identical to the first impact as a further aid in identifying the particular effects of the second impact.\par
\begin{figure}
\centerline{\includegraphics[angle=-90,viewport=165 50 530 814,clip,width=\textwidth]{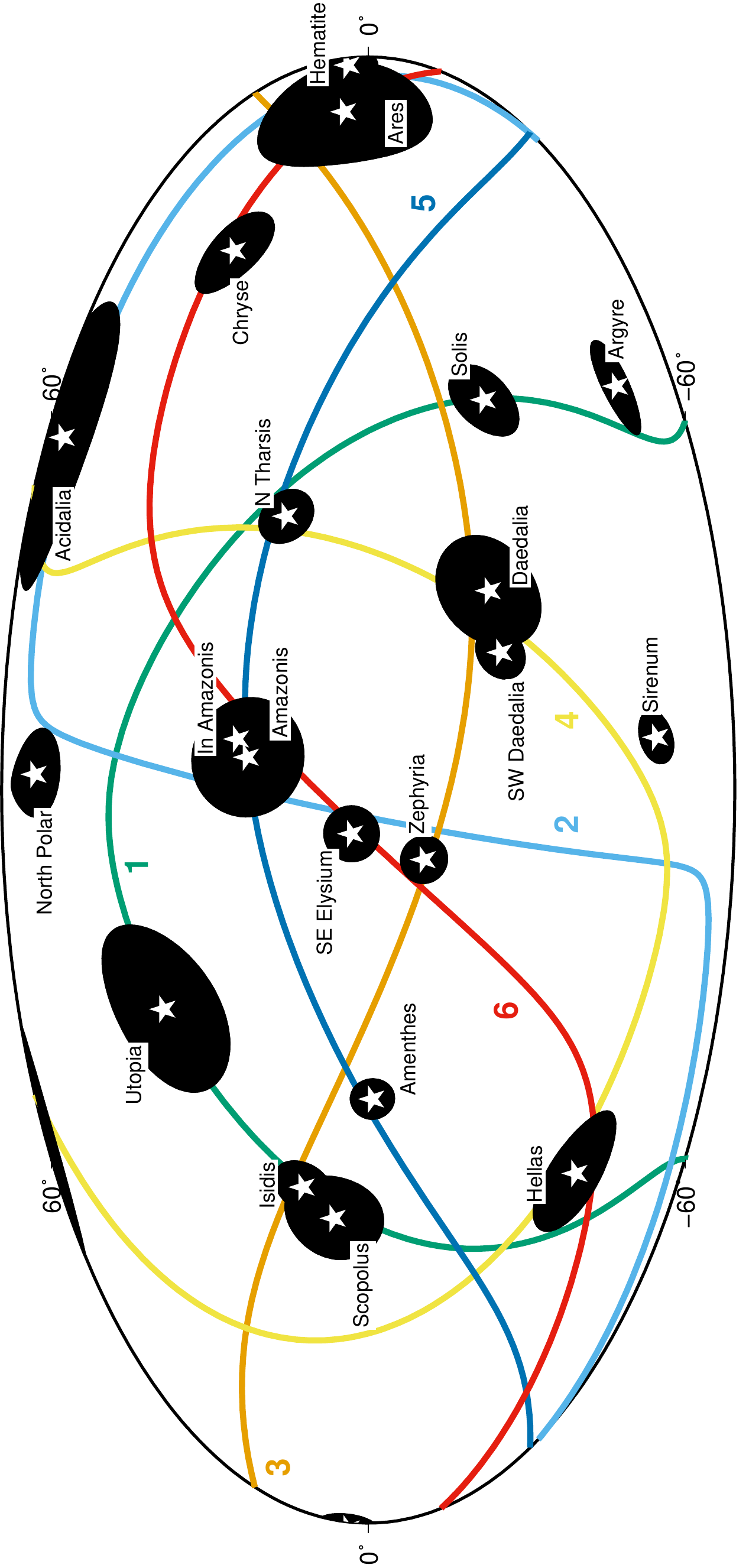}}
\caption{Schematic map in Hammer projection of the locations of the large impact basins on Mars after \citet{Frey08} listed in Table~\ref{tab:impacts} and the great circles (colored curves) by which they are assigned to 2D models.\label{fig:gcmap}}
\end{figure}

The second subset of models includes six models with a succession of four to eight large impacts on Mars taken from the list of 20 large basins by \citet{Frey08} \citep[cf.][]{JHRobe:etal09}, whereby the impact sequences are chosen such that all events in a given model lie approximately on a great circle (GC) (Fig.~\ref{fig:gcmap}). The spacing of the impacts corresponds to the approximate position of the basin centers on the great circle, and their timing follows the ages given by \citet{Frey08}; the method of mapping them is outlined in \ref{app:gc}. This model set thus combines a certain element of randomness with the constraints from observations on Mars.
\begin{sidewaystable}
\centering
\caption{Impact basins considered in this study in order of descending age, after \citet{Frey08} and \citet{JHRobe:etal09}. The impactor diameter $D_\mathrm{imp}$ and the diameter $D_\mathrm{ic}$ and depth $z_\mathrm{ic}$ of the isobaric core are estimates from scaling laws derived from the observed final basin diameter $D_\mathrm{f}$. The ages and positions in the list apply only to impacts in the GC models; the position given in the ``Position'' column are the geographical coordinates, the single degree entries in the GC columns are the positions on the great circle of the GC model determined as described in \ref{app:gc}, measured clockwise from the bottom of the circle in the cross-section images.\label{tab:impacts}}
\begin{tabular}{lccccccccccccc}\toprule
Basin/Code&Age&$D_\mathrm{imp}$&$D_\mathrm{ic}$&$z_\mathrm{ic}$&$D_\mathrm{f}$&Position&\multicolumn{7}{c}{Model sets}\\
&(Myr)&(km)&(km)&(km)&(km)&&GC1&GC2&GC3&GC4&GC5&GC6&U/I-*\\\midrule
Amenthes (Ame)&4218&188.3&103.8&51.8&1070&110.6\textdegree E/0.9\textdegree S&&&73.0\textdegree&&271.5\textdegree&&\\
Zephyria (Z)&4210&213.2&117.4&58.6&1193&164.3\textdegree E/12.4\textdegree S&&255.9\textdegree&126.9\textdegree&&&257.6\textdegree&\\
Daedalia (Dae)&4199&526.9&290.4&144.0&2639&228.3\textdegree E/26.5\textdegree S&&&188.1\textdegree&243.4\textdegree&&&\\
Sirenum (Sir)&4196&188.0&103.6&51.8&1069&205.3\textdegree E/67.4\textdegree S&&&&200.0\textdegree&&&\\
SW Daedalia (SWD)&4176&231.9&127.8&63.3&1278&213.9\textdegree E/29.4\textdegree S&&&175.3\textdegree&235.6\textdegree&&&\\
Ares (Are)&4167&680.5&375.0&185.4&3300&343.9\textdegree E/4.0\textdegree N&&84.0\textdegree&303.8\textdegree&&&83.5\textdegree&\\
Amazonis (Ama)&4154&580.9&320.2&158.5&2873&187.9\textdegree E/27.1\textdegree N&&299.5\textdegree&&&350.7\textdegree&302.1\textdegree&\\
In Amazonis (IA)&4152&207.2&114.2&56.5&1156&192.5\textdegree E/29.3\textdegree N&&302.9\textdegree&&&354.8\textdegree&306.6\textdegree&\\
Solis (Sol)&4148&311.1&171.4&85.4&1663&275.3\textdegree E/23.8\textdegree S&117.5\textdegree&&230.4\textdegree&&&&\\
N Tharsis (NTh)&4143&244.8&134.8&67.3&1347&243.6\textdegree E/17.6\textdegree N&66.0\textdegree&&&290.0\textdegree&42.6\textdegree&&\\
Chryse (C)&4140&324.3&178.6&89.0&1725&318.0\textdegree E/25.0\textdegree N&&&&&&51.0\textdegree&\\
Hematite (Hem)&4138&189.1&104.2&51.5&1065&357.8\textdegree E/3.2\textdegree N&&87.6\textdegree&&&&93.5\textdegree&\\
Scopolus (Sc)&4133&439.5&242.2&120.2&2250&81.8\textdegree E/6.9\textdegree N&276.4\textdegree&&42.9\textdegree&&&&\\
Acidalia (Aci)&4132&629.6&347.1&172.0&3087&342.7\textdegree E/59.8\textdegree N&&29.7\textdegree&&9.6\textdegree&&&\\
North Polar (NP)&4124&297.7&164.0&81.8&1600&195.2\textdegree E/80.0\textdegree N&&350.6\textdegree&&&&&\\
Utopia (U)&4111&699.5&385.4&190.5&3380&115.5\textdegree E/45.0\textdegree N&324.3\textdegree&&&&&&(\checkmark)\\
SE Elysium (SEE)&4107&256.2&141.1&70.5&1403&170.3\textdegree E/3.7\textdegree N&&272.9\textdegree&&&&273.5\textdegree&\\
Hellas (Hel)&4065&399.5&220.1&109.4&2070&66.4\textdegree E/42.3\textdegree S&226.6\textdegree&&&133.7\textdegree&&170.6\textdegree&\\
Argyre (Arg)&4043&238.2&131.2&65.5&1315&317.5\textdegree E/49.0\textdegree S&158.6\textdegree&&&&&&\\
Isidis (I)&3810&247.5&136.4&67.5&1352&87.8\textdegree E/13.4\textdegree N&284.9\textdegree&&46.8\textdegree&&&&(\checkmark)\\\bottomrule
\end{tabular}
\end{sidewaystable}

\section{Results}
All models in this study are derived from an impact-free reference model whose setup is identical with the 36\,ppm H$_2$O reference models of \citet{RuBr17c,RuBr18b} except that it starts at 4.5\,Ga instead of 4.4\,Ga. The reference model starts with a regular pattern of upwellings and downwellings that transitions into a more chaotic mode after a few tens of millions of years. After 500--600 million years, the flow reorganizes itself into a more regular large-scale pattern again over the course of several hundreds of millions of years, ending up with a long-lived final configuration with five approximately regularly spaced large plumes while cooling steadily and forming a thick thermal boundary layer at the surface. The initially superheated core cools quite rapidly and loses its excess heat in the first few hundreds of millions of years, but it never begins to crystallize. The divergence of the evolutionary paths of the models thus begins with the first impact striking in a given model. For the two-impact models, this occurs early in the transition from the chaotic to the final quasi-static configuration, for the great-circle models it starts at some point during the chaotic stage. Hence, all impacts, and especially the earliest ones, strike a planet with a still thin lithosphere and hot, globally vigorously convecting mantle whose uppermost part is undergoing some regular global melting, albeit at quite low intensity.
\subsection{Two-impact models}
\begin{figure}
\includegraphics[width=\textwidth]{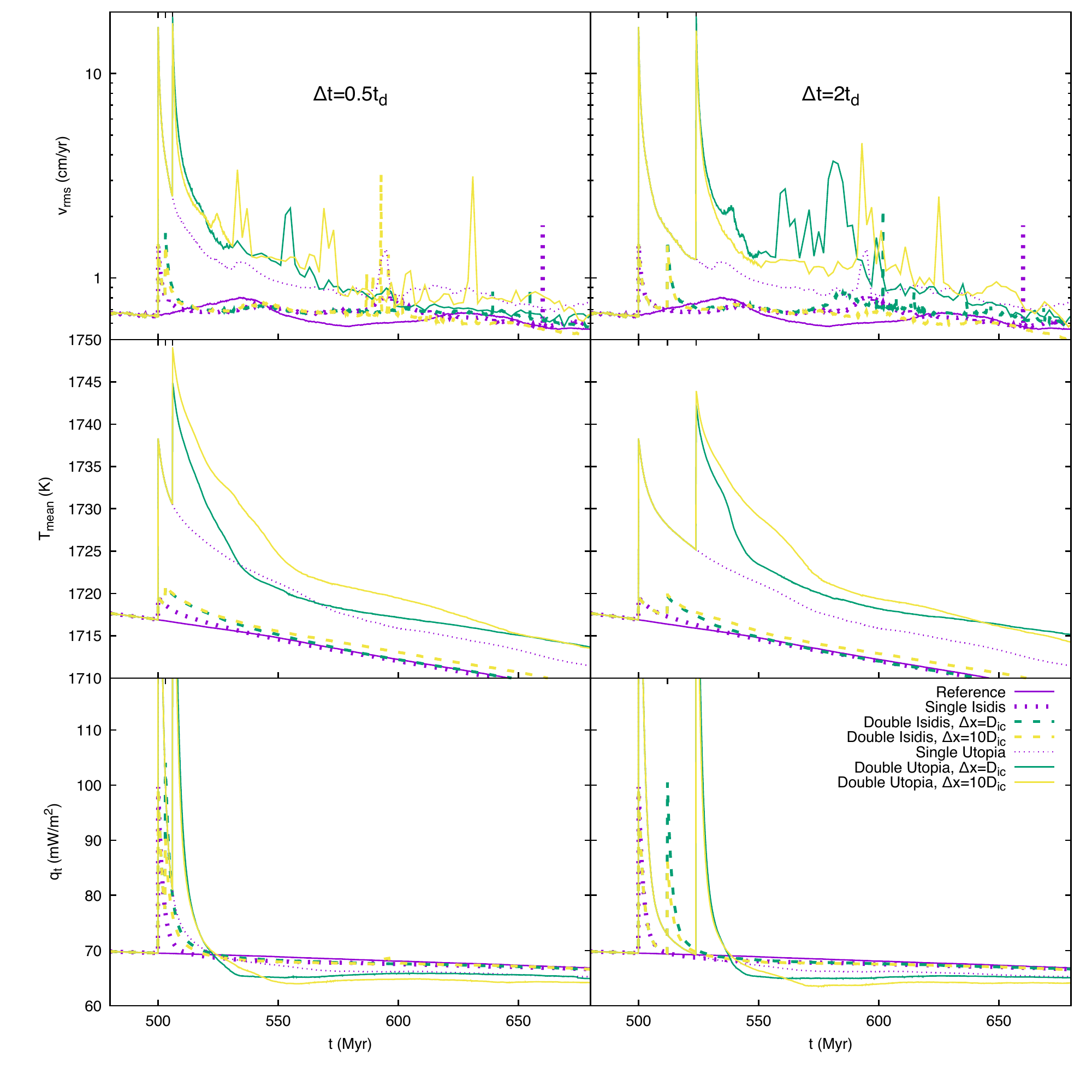}
\caption{Root-mean-square velocity, mean temperature, and surface heat flow for selected models with two impacts of either Isidis-like or Utopia-like magnitude. The impact-free and the single-impact models are also shown for reference. The Utopia models have been smoothed at $t>530$--550\,My by averaging over 2\,My-intervals for better readability. The ticmarks on the top abscissas mark the times at which impacts occurred in any of the models. A collection of plots for all models can be found in the Supplementary Material.\label{fig:vTq-t-IU}}
\end{figure}

The quasi-instantaneous input of energy into the planetary interior by an impact produces a steep change in the temporal evolution of several variables that serve as indicators of the dynamical state of the mantle, e.g., the global means of the root-mean-square velocity, $v_\mathrm{rms}$, the mantle temperature, $T_\mathrm{mean}$, and the heat flows through the surface and the CMB, $q_\mathrm{t}$, and $q_\mathrm{CMB}$. This step-like change is followed by an initially rapid rebound towards the pre-impact state over the next few millions of years (Fig.~\ref{fig:vTq-t-IU}). The signal of the second impact is added to the decaying signature of the first in different ways for different variables. To compare the second impact with the first, we determine the maximum change of these variables due to each impact and calculate the ratio of the second signal to the first; e.g., for $T_\mathrm{mean}$ the second-to-first ratio of impact effect amplitudes would be
\begin{equation}
\alpha_T=\frac{T_{\mathrm{mean}}(t_2)-T_{\mathrm{mean,sngl}}(t_2)}{T_{\mathrm{mean}}(t_1)-T_{\mathrm{mean,ref}}(t_1)},
\end{equation}
where ``sngl'' and ``ref'' indicate the corresponding models with a single impact and the impact-free reference model, respectively, and $t_1$ and $t_2$ indicate the times of the first and the second impact. These ratios are expressed as functions of the normalized distance, $\Delta x'=\Delta x/D_\mathrm{ic}$, and time lapse, $\Delta t'=\Delta t/t_\mathrm{d}$, respectively. If this ratio is smaller than 1, i.e., if the change of the indicator variable is smaller in the second impact than in the first, that would suggest that the effect is not simply additive but may tend towards a state of saturation. If the ratio were equal to 1, impacts with the same parameters would have identical effects, implying that the target has no memory of the effects of preceding impacts. $\alpha_X$ is therefore a measure of the superposition efficiency of impacts with respect to the indicator variable $X$.\par
\begin{figure}
\centering\includegraphics[width=\textwidth]{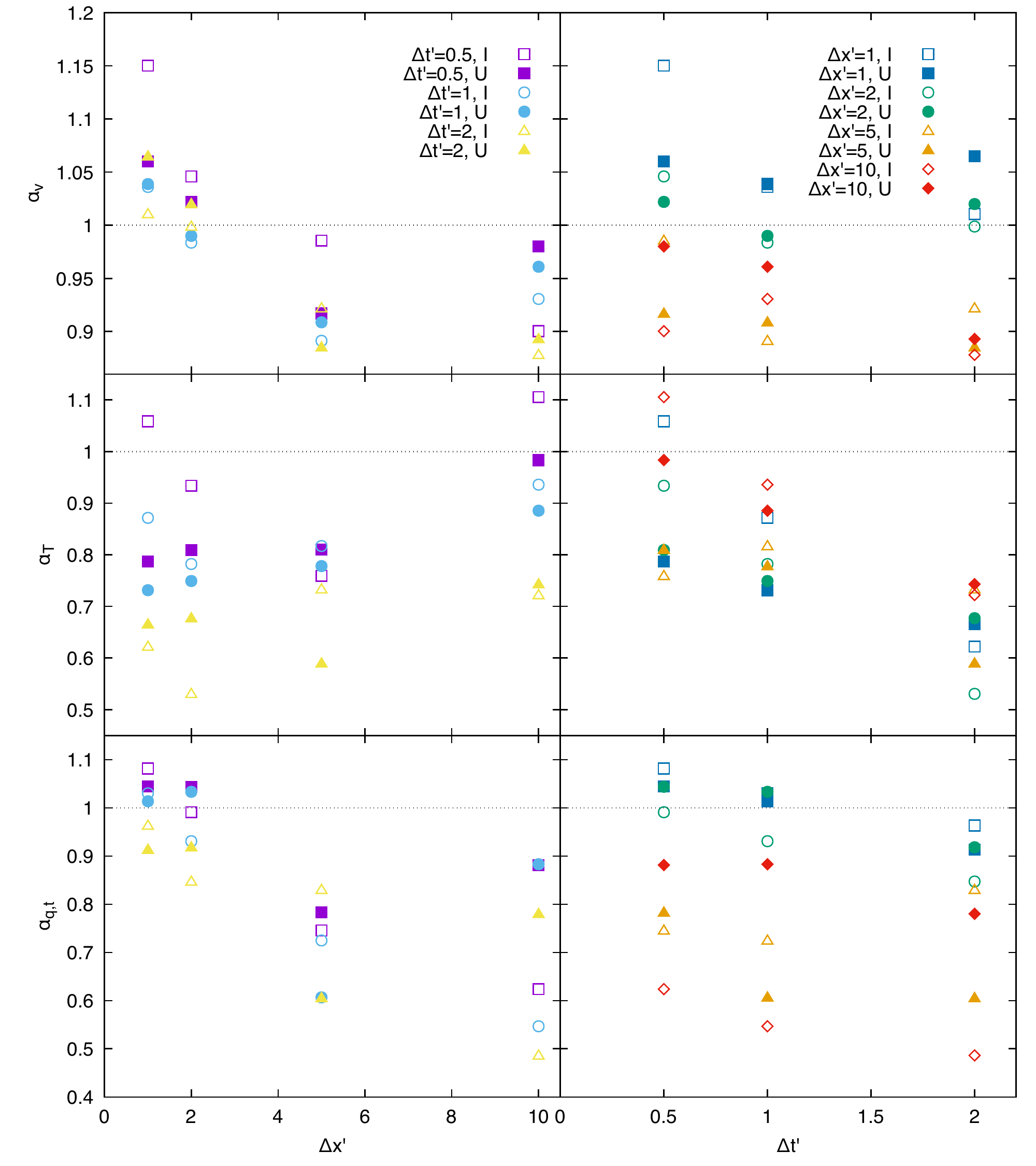}
\caption{Second-to-first ratio of impact effect amplitudes (superposition efficiencies) of $v_\mathrm{rms}$, $T_\mathrm{mean}$, and $q_\mathrm{t}$ as functions of $\Delta x'$ (left column) and $\Delta t'$ (right column).\label{fig:vTq}}
\end{figure}

The models show that the superposition efficiency for $v_\mathrm{rms}(t)$ tends to increase with decreasing $\Delta x'$ and also, although much less clearly, with $\Delta t'$ (Fig.~\ref{fig:vTq}), because the lingering thermal anomaly from the first impact boosts the upwelling triggered by the second. The physical cause for this boost may be that in closely spaced anomalies whose temperatures decrease away from the central axis, the superposition leads to particularly high peak temperatures near the axis, which have a non-linear positive effect on the viscosity and thus on flow velocities. A simplified analytical model for pipe-like flow with temperature-dependent viscosity indicates that a strong focusing of the flow and a speedup by more than an order of magnitude relative to the isoviscous case is possible for peak temperature anomalies of a few hundred kelvins (cf.~\ref{app:pois}). $\alpha_v(\Delta x',\Delta t')$ is not linear in both variables, and the data suggest that there may be a slight increase beyond a certain $\Delta x$ in the Utopia-size models, which may be due to the increase of the affected total volume. Except for closely spaced impact pairs, $\alpha_v<1$. Later peaks in $v_\mathrm{rms}(t)$ in Fig.~\ref{fig:vTq-t-IU} are not directly related to the impacts and reflect other mantle dynamical processes such as rising plumes or delamination of crust. They do not appear in the impact-free reference model and are an indirect, longer-term consequence of local lithospheric instabilities induced by the impacts.\par
To assess the thermal effect, we consider first a simplified setup consisting of two identical spherical anomalies of radius $R$ whose centers are separated by $\Delta x$; for $\Delta x<2R$, these anomalies would overlap to some extent. Especially for impacts in short succession, the thermal anomaly from the first strike is still so strong that the joint anomaly after the second one can be treated as a homogeneous volume as an approximation. The magnitude of the thermal anomaly formed by overlapping spheres is proportional to its volume $V(\Delta x)$, and the functional form of $V$ points to a sublinear increase with $\Delta x$ up to $\Delta x \leq 2R$ (cf.~\ref{app:anoVA}); for larger offsets, it should be constant, as both anomalies are independent from each other. Estimating the cooling time of a spherical anomaly from the analytical solution for diffusive cooling of a solid sphere, one finds that the temporal decay is governed by a term of the form $\exp(-\pi^2\kappa t/R^2)$, for a diffusivity $\kappa$ \citep[e.g.,][]{Yeom15}. The level from which the mantle is reheated by the second impact depends on how much the first anomaly has cooled by the time it strikes, and so we expect a decrease of the thermal superposition efficiency $\alpha_T$ that approximately reproduces that trend, whereby the effective diffusivity is enhanced by convection within the anomaly and entrainment of cooler material from outside. The trends in Fig.~\ref{fig:vTq} confirm that $\alpha_T$ grows approximately linearly with spatial separation and decreases approximately linearly with temporal separation; the apparently linear form of the latter is due to the limited range of $\Delta t'$ considered, whose lowest value of 0.5 lies already after the stage of strongest cooling. The increase with $\Delta x'$ is clearer for the Utopia-size impacts, presumably because they penetrate more deeply into the sublithospheric mantle and are hence less affected by the cold lid. As $\alpha_T<1$ in almost all of the two-impact models, the mantles move towards a state of saturation in which further impacts ideally would not entail more heating but transform their energy otherwise, e.g., by melt production.\par
The signature of an impact on the global surface heat flow is proportional to the surface area affected, whereby the strongest effects are expected within the final crater. Under the assumptions made in our models, we expect that the signatures of multiple impacts are only additive if the distance between two impacts exceeds a certain value, because otherwise the affected areas will overlap to some extent. Impact modification of a surface region that has already been modified by a previous impact is reduced, because processes like pore filling of regolith by volcanism can only occur once. In general, the superposition efficiency $\alpha_{q,\mathrm{t}}$ decreases approximately linearly with increasing spatial and temporal separation (see Fig.~\ref{fig:vTq} and \ref{app:qcrat}); however, for the large Utopia impacts, there is a clear minimum at $\Delta x'=5$ at all $\Delta t'$, which indicates that other processes also leave their mark on $q_\mathrm{t}(t)$. The CMB heat flow, by contrast, shows only minute variations as a direct consequence of the impact in our models, and no meaningful dependence of the peak amplitudes on $\Delta x'$ or $\Delta t'$ is found (cf. Fig.~\ref{fig:Score}, left column).\par
\begin{figure}
\includegraphics[width=\textwidth]{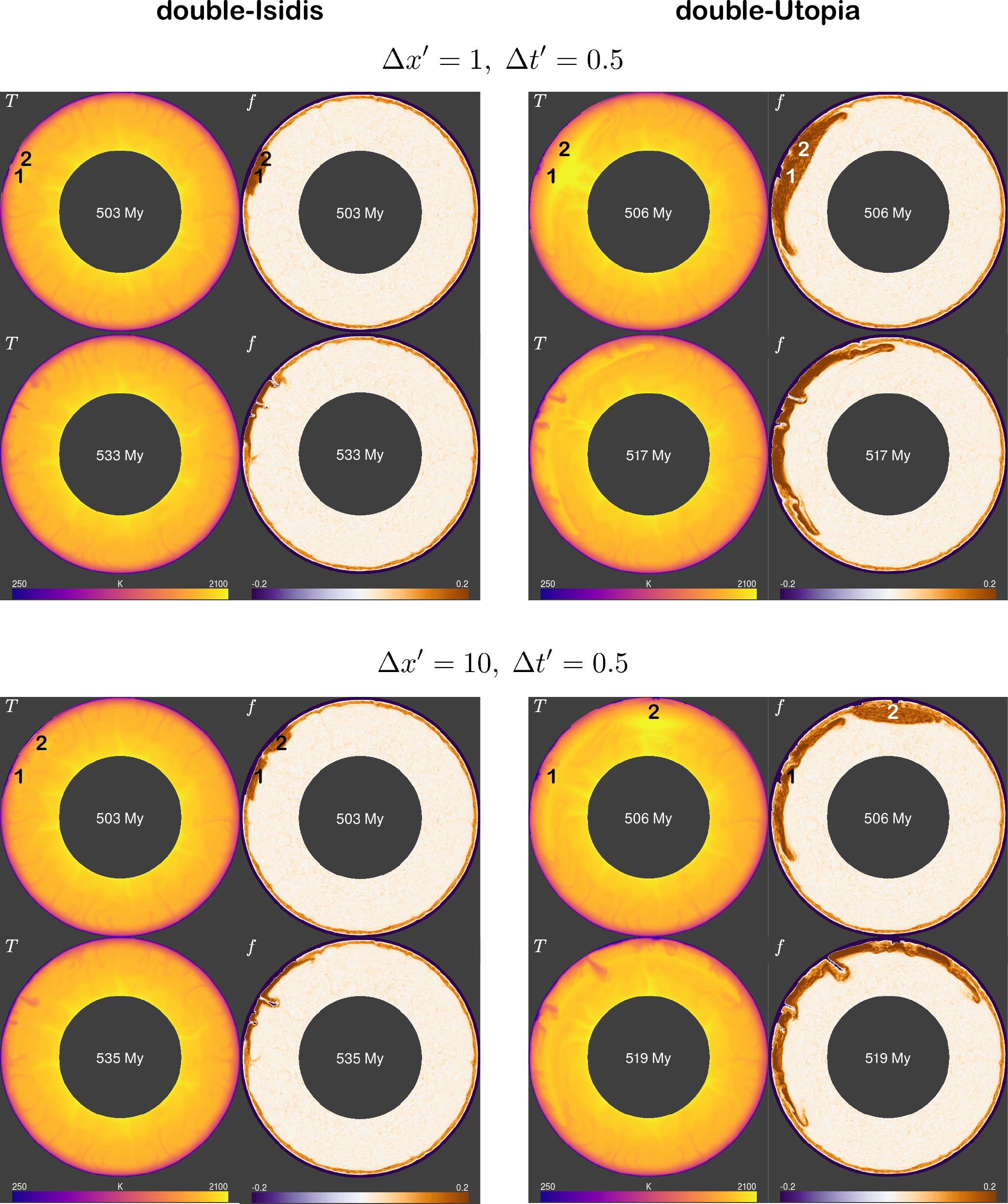}
\caption{Snapshots of temperature ($T$, left column) and depletion ($f$, right column) for the models with the closest and widest impact spacing and the shortest time interval. The impact locations are marked in the upper rows of each block as \textsf{1} and \textsf{2}. The first impact occurs at 500\,My in all models, the second impact follows at 503\,My in the double-Isidis models and at 506\,My in the double-Utopia models, respectively (upper rows of each block). The evolution immediately following the second impact is shown in the snapshots in the lower rows of each block.\label{fig:snap-IU}}
\end{figure}

The shock-heated volumes are also depleted in fusible components and are thus compositionally distinct and less dense than pristine mantle material, which contributes to their buoyancy and the reinforcement of convection caused by impacts. On the other hand, a mantle volume that was strongly depleted by extensive melting in the first impact cannot experience much further depletion by melting from a second nearby impact. Thus, the buoyancy of the flow after the latter should be expected to be more thermally dominated in models with small $\Delta x'$, but the potentially diminished compositional contribution to the total buoyancy is apparently not significant enough to compensate the other reinforcing effects on $v_\mathrm{rms}$ discussed above. As the impact-generated anomalies ascend and spread out beneath the lid, they may not only influence each other dynamically, as seen in the $v_\mathrm{rms}(t)$ curves, but will also collide and merge, especially if they are closely spaced. In models with widely spaced impacts in close succession, both anomalies develop somewhat independently for a certain time until they have spread far enough below the lid to run into each other. In such cases a piece of normal mantle can get caught between them and induce a downwelling due to its relatively higher density, especially for smaller impacts with less vigorous dynamics. An example for this can be seen in the double-Utopia model with $\Delta x'=10$ and $\Delta t'=0.5$ in the lower right block of Fig.~\ref{fig:snap-IU}.\par
\begin{figure}
\includegraphics[viewport=0 140 1080 1440,clip,width=\textwidth]{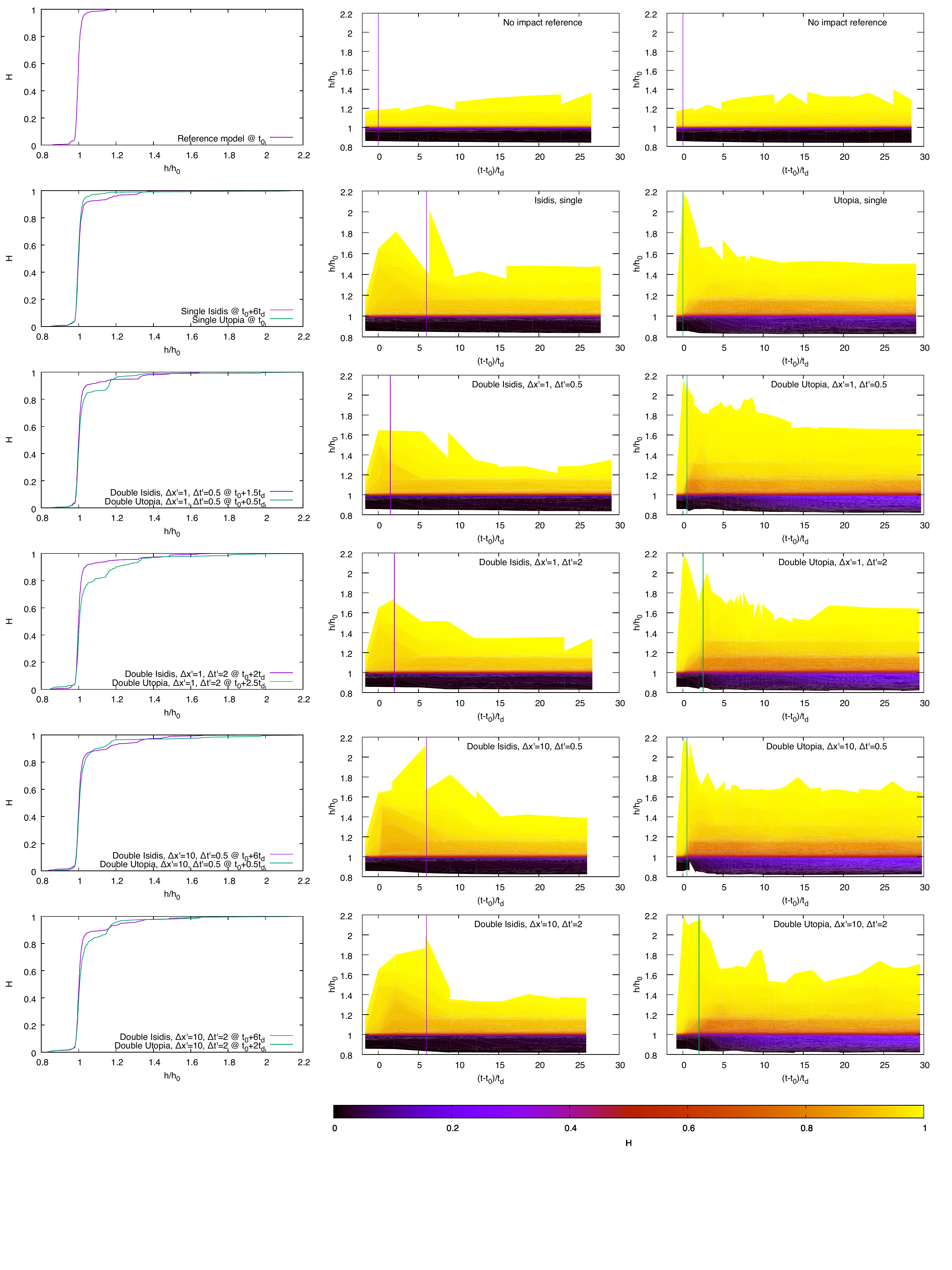}
\caption{Cumulative distribution function $H$ of the crustal thickness of the two-impact model series from Fig.~\ref{fig:vTq-t-IU} as a function of crustal thickness $h$ and time $t$ at and after the impacts. The left column shows $H(h)$ at that time after the second impact at which the crust reaches its maximum thickness. These curves are thus sections of the $H(h,t)$ maps in the other two columns parallel to the ordinate; they are marked there as vertical lines in the corresponding color. The crustal thickness is normalized with the mean crustal thickness $h_0$ immediately before the first impact, i.e., at $t_0=500$\,Myr, which can be calculated from the impact-free reference model. The abscissa in the middle and right column gives the time since the first impact in multiples of the decay timescale $t_\mathrm{d}$, which is $\sim 6$\,Myr for the Isidis-sized impacts (left column) and $\sim 12$\,Myr for the Utopia-sized impacts (right column). The impact-free and the single-impact models are also shown for comparison. The full set of figures is provided in the Supplementary Material.\label{fig:hcumt-IU}}
\end{figure}

\begin{figure}
\includegraphics[width=\textwidth]{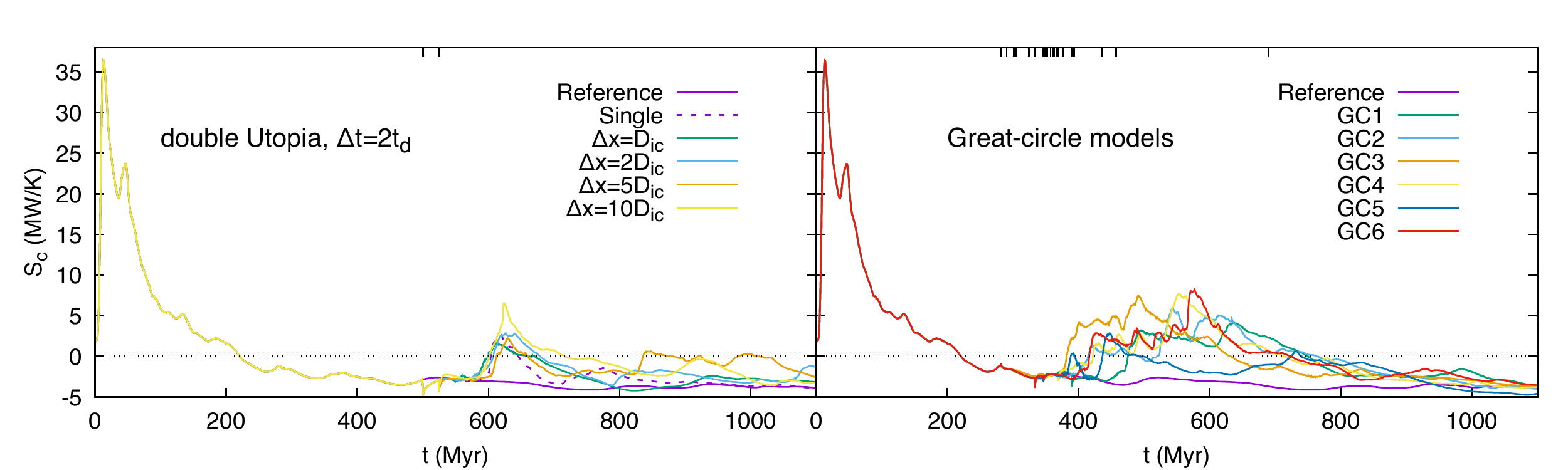}
\caption{Total entropy production of the core for the double-Utopia models with the longest time interval between impacts (left) and for the great-circle models (right); a dynamo can exist if $S_\mathrm{c}>0$. The corresponding heat flows at the CMB can be found in the Supplementary Material. The ticmarks on the top abscissa mark the times at which impacts occurred in any of the models.\label{fig:Score}}
\end{figure}

Heat and enhanced convection can also result in increased production of melt and crust, but the extreme depletion of the shallower mantle in the impact-affected region counteracts this effect. Global crustal thickness evolution maps $h(x,t)$ and the corresponding cumulative distributions $H(h,t)$ (Figs.~\ref{fig:hcumt-IU}, S3--6) show that impacts broaden the range of thickness by creating thinner and thicker regions, especially in the large (Utopia-sized) impact model series. In the following, we will normalize $h$ with the mean crustal thickness $h_0$ at the time $t_0$ of the first impact, and we express the time elapsed since $t_0$ in multiples of the decay time $t_\mathrm{d}$ in order to achieve a scale-free representation that might reveal common trends for impacts of different magnitudes. The cumulative distribution function of the crustal thickness $H(h,t)$ quantifies how large the fraction of crust with a thickness no larger than $h$ is at the time $t$; for example, in the scale-free plots of some selected cases in Fig.~\ref{fig:hcumt-IU}, $H(1.1, t_0+0.5t_\mathrm{d})=0.87$ means that at half a decay time after the first impact, 87\% of the entire crust has a thickness of $1.1h_0$ or less. The curves in the left column of the figure show $H(h)$ at that time after the second impact at which a part of the crust reaches near-maximum thickness and has a diverse thickness distribution. As the curve for the reference model shows, the thickness distribution in general is essentially unimodal and shows little scatter around the mean value $h/h_0=1$, i.e., the crust has a rather uniform thickness. The unimodality remains largely valid in the models with impacts as well, but the steep increase at 1 is not quite as high, whereas the flanks of the curve become longer and higher, reflecting the broadening of the thickness distribution due to production of anomalous crust by impacts. The color-coded maps in the middle and right column of Fig.~\ref{fig:hcumt-IU} show how this cumulative distribution evolves through time in a time window around (mostly after) the impact time; the curves in the left column are thus sections of these maps in the vertical direction and are marked as vertical lines in the map plots. The spatial pattern and evolution of post-impact crustal thickness depends on spacing, timing, and size of the impacts. Comparison of the impact-free and the single-impact model shows clearly that the additional melt extraction after the impact results in a marked and lasting increase in $h$ in some regions, especially near the site of a Utopia-sized impact (cf. Fig.~S3). In the Utopia series, the second impact then increases the fraction of thickened crust, possibly including some more growth of already thickened regions, but at least in models in which the impacts are further apart in space and time, the most extreme effect is not as long-lived. On the other hand, some parts of the crust become thinner a certain time after the impacts (usually $5t_\mathrm{d}$ to $15t_\mathrm{d}$ after $t_0$), which seems to be related to delamination of very thick crustal roots, as the increase in the fraction of thinner-than-average crust is related to the gradual disappearance of some of the thickest crust in the $H$ diagram. Note that the rather discrete, step-like thickness increments are partly an artifact of the grid discretization, which limits the accuracy with which the crustal thickness can be determined. In the Isidis series, these features are less clear, and the thinning is much less significant as less delamination takes place. The occurrence of extra-thick crust due to the second impact, however, is also visible but has a much more transient character and fairly uniform lifetime of about $10t_\mathrm{d}$ in the entire parameter range covered. As a consequence of the interaction of the two anomalies, the location of the thickest post-impact crust does not necessarily coincide exactly with the impact site in both series.\par
In all cases investigated, the differences between the models diminish with time, and the impact signature in the temporal evolution of the system's dynamical variables fades and has disappeared long before the present. In particular, thermal anomalies fade with time and leave no signal in the present-day heat flux. Compositional anomalies, however, are preserved, as pointed out by \cite{RuBr17c}, but it would be difficult to draw a sharp boundary between the region of influence of one impact or the other if the impacts are close enough to overlap. Variations in post-impact crust formation are also reflected in the crustal thickness and may thus preserve a long-term record of impact-induced mantle dynamics.\par
In addition to the effects of impacts at the surface or in the shallow mantle, it has been proposed that impacts also affect a core dynamo that may be active \citep[e.g.,][]{JHRoAr-Ha14}. We did not include any direct impact effects such as shock heating of the core in our treatment as did those authors and therefore neglect the potential effects of an impact-generated hot layer at the top of the core whose appearance was proposed by them. However, given that even the largest impacts in our models are somewhat smaller than the one modeled by \citet{JHRoAr-Ha14} and that shock heating decreases to nearly zero towards the CMB in our models, we do not expect a significant direct effect of the shock on the core. Impact-related variations of the heat flow through the CMB would therefore mostly be induced by post-impact mantle processes. The left panel of Fig.~\ref{fig:Score} shows the total entropy production of the core, which is a good proxy for $q_\mathrm{CMB}(t)$ in our models, for the first billion years in the double-Utopia models with the longest time interval between impacts ($\Delta t'=2$); the curves for smaller $\Delta t'$ look very similar, whereas in the double-Isidis models the deviations from the reference are negligible. At the times of the impacts, there are tiny downward excursions that indicate a short period of reduced entropy production and heat flow out of the core during the presence of the thermal anomaly in the mantle; this effect would be compensated if shock heating also affected the core. The more significant effect, however, sets in only several decay times after the second impact and consists in an increased entropy production and heat flow out of the core. It is caused by the arrival at the CMB of cooler lithospheric material into which small fragments of eclogitized crust are embedded and that had become unstable in the aftermath of the impacts, delaminated, and sunk to the bottom of the mantle where it cools the core from above. However, in the lower part of the martian mantle, eclogite has a significantly smaller density excess than in the shallow mantle \citep[e.g.,][]{AoTa04}, and as the crustal fragments form a close union with the lighter harzburgitic residue that went down along with them, the old lithosphere does not accumulate permanently at the CMB. Instead, it is heated enough by its internal sources in the eclogite and by conduction from the surrounding warm mantle to become unstable and be mixed back into the mantle again by the general mantle flow. Hence, while spacing and time interval between impacts are unimportant for post-impact effects on the CMB, the size and, to some degree, the number of the impacts are decisive, because they control the extent of lithospheric instability.

\subsection{Great-circle models}
\begin{figure}
\includegraphics[width=\textwidth]{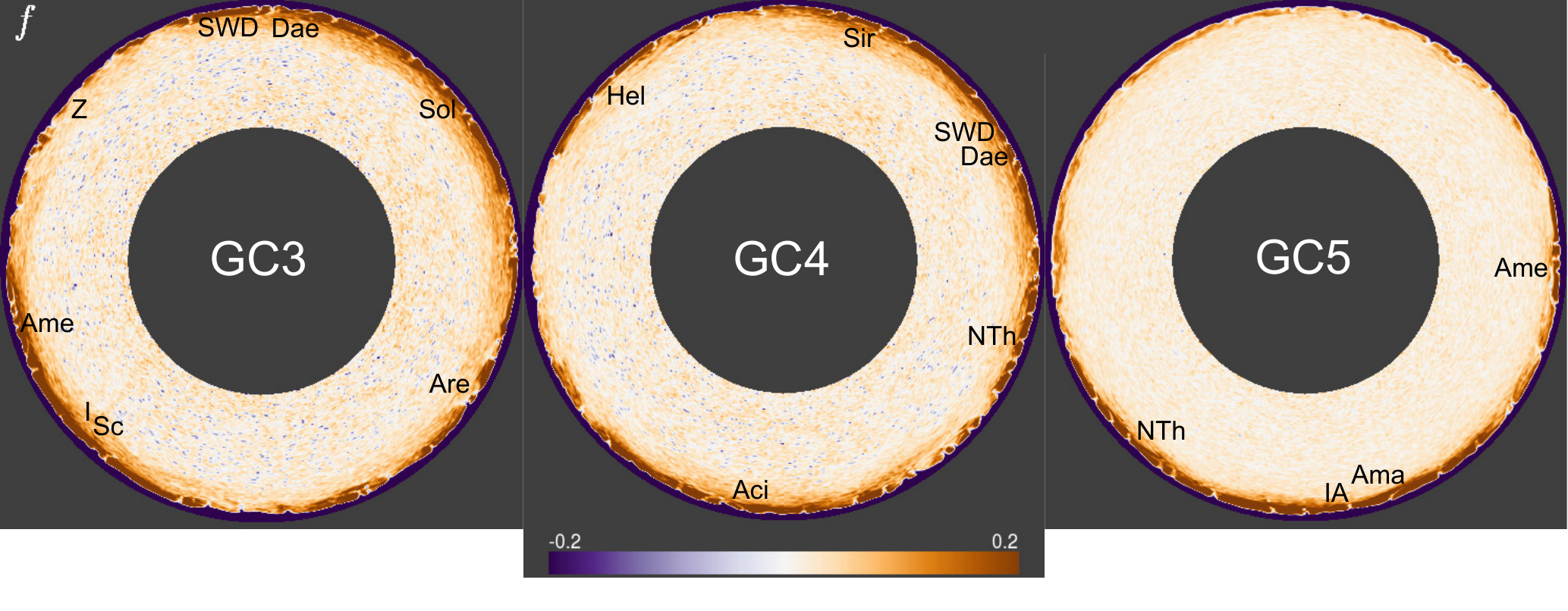}
\caption{Snapshots of the present-day depletion patterns for the great-circle models 3, 4, and 5. The northernmost points of the respective great circles lie at the six o'clock position. Snapshots of all models are presented in the Supplementary Material.\label{fig:cnow}}
\end{figure}

As the models with multiple impacts on a great circle show, a succession of various different large impacts produces a strongly variable depletion pattern in the uppermost mantle (Fig.~\ref{fig:cnow}). This pattern still reflects the diversity of the impacts that produce it, but the vigorous post-impact dynamics and merging of the individual anomalies precludes the distinction of clear boundaries between the traces of the discrete events. As the individual impacts in these models lie further apart from each other and span a longer time interval than those in the previous sets, the more indirect effects as a whole, especially those on the lithosphere, also last longer and extend to the entire planet: specifically, crustal delamination as a long-term consequence of the impacts becomes a planet-wide process and lasts much longer than in the previous sets.\par
\begin{figure}
\includegraphics[height=0.9\textheight]{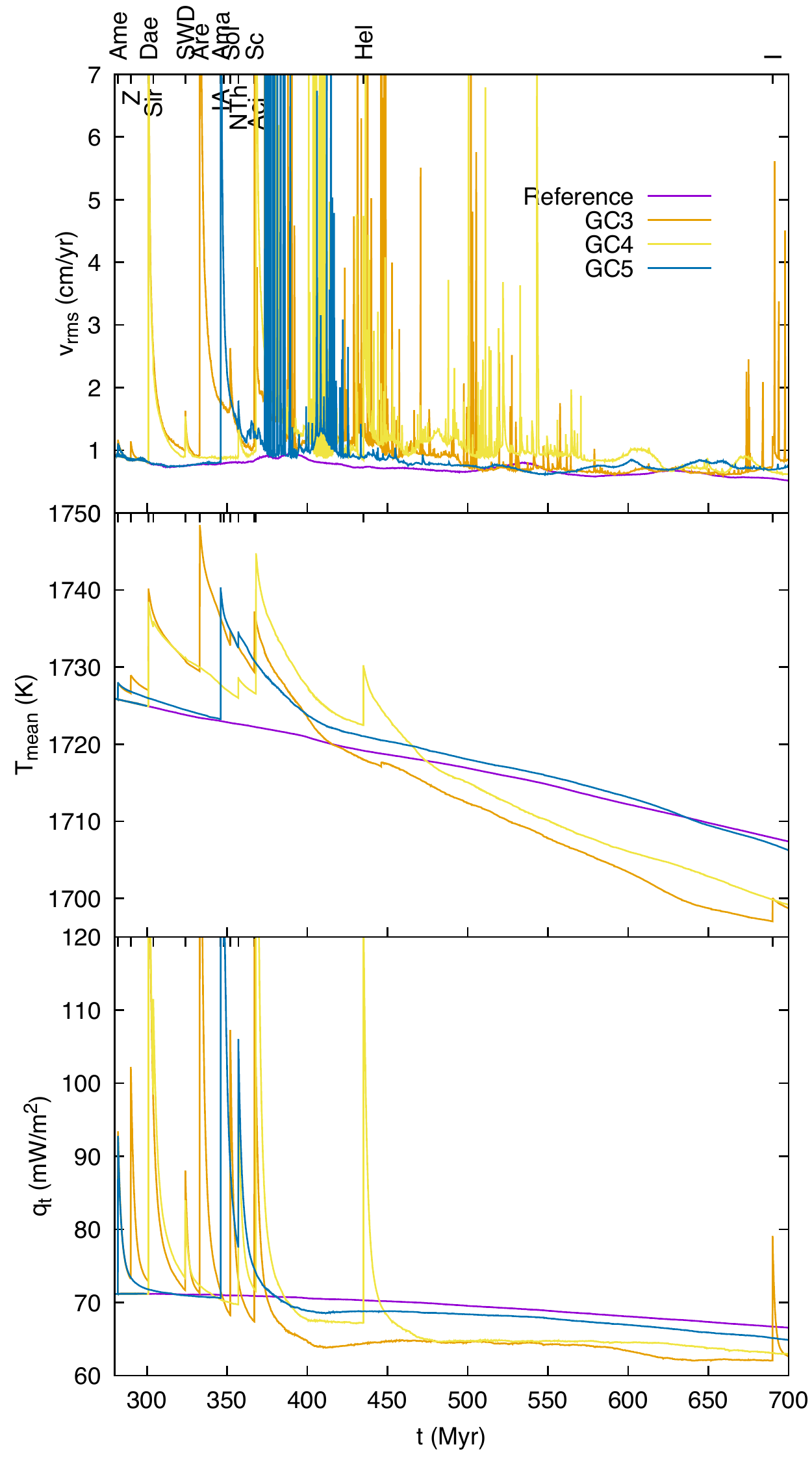}
\caption{Root-mean-square velocity, mean temperature, and surface heat flow for the great-circle models 3, 4, and 5, which include 8, 6, and 4 impacts of varying magnitudes, respectively. The impact-free model is also shown for reference. A collection of plots for all models can be found in the Supplementary Material.\label{fig:vTq-t-gc345}}
\end{figure}

Figure~\ref{fig:vTq-t-gc345} shows the root-mean-square convection flow velocity, the mantle mean temperature, and the surface heat flow for models GC3, GC4, and GC5, which represent the models with the highest, an intermediate, and the lowest number of impacts on a great circle in the model set. The basic shape of the curves resembles those from the idealized double-impact models in the previous section. The impacts are again marked by sharp peaks with subsequent gradual decays of these variables; later narrow peaks in $v_\mathrm{rms}$ correspond to the quick motion of cold lithospheric instabilities formed in the aftermath of the impacts, mostly related to the formation of thick crust. Although we do not normalize times here with the decay time $t_\mathrm{d}$ because there is no single $t_\mathrm{d}$ applicable to the entire duration of a given model, it is worth noting that estimates for the $t_\mathrm{d}$ for individual impacts seem to follow the aforementioned relation between $t_\mathrm{d}$ and the size of the thermal anomaly, $\exp(-\pi^2\kappa t/R^2)$. The center panel of Figure~\ref{fig:vTq-t-gc345} shows that the $T_\mathrm{mean}$ of models with particularly intense impact activity even drops below the $T_\mathrm{mean}$ of the impact-free reference model some time after the impacts. The reason for this maybe unexpected behavior is that the heat from the impact shocks has largely diffused away after some tens of millions of years, but the copious amounts of cool crust that were produced in their aftermath are becoming unstable and partly sink into the mantle, where they form efficient heat sinks; the crust is so cold, because the melt from which it is formed has been set to the surface temperature upon eruption. At some point between 1.5\,Gy and 2.5\,Gy, i.e., outside the plotted range, all of the models return to and remain at $T_\mathrm{mean}$ values above the reference as the delaminated crust has been warmed up by its own internal radioactive heat production. In all curves, the largest impacts dominate strongly the total deviation from the impact-free reference. In the Introduction, we had speculated that at some point, a planetary mantle could reach a ``saturated'' state in the sense that additional energy from another impact does not reinvigorate the dynamics, for instance because the shallow mantle is so hot and/or depleted throughout that some sort of stable layering is established that suppresses further whole-mantle convection. However, in none of the models do we find any clear indication of such a ``saturation'', which suggests that Mars was always far enough away from it, if such a state is possible at all. Nonetheless, and even though the larger spatial and temporal separation of the individual events diminishes the direct interactions between impacts, the different histories of the GC models induce different dynamical states of the mantle that lead to different signatures between models for a given impact. This becomes most readily visible in the two late impacts, Hellas (GC1, GC4, GC6) and Isidis (GC1, GC3). Both events leave signals of different magnitude in the different models in which they occur as a consequence of the different evolutionary paths (Fig.~S7). In all six models, however, the global characteristics return to a state close to that of the reference on timescales that are substantially shorter than the age of the planet.\par
\begin{figure}
\includegraphics[viewport=0 115 720 864,clip,width=\textwidth]{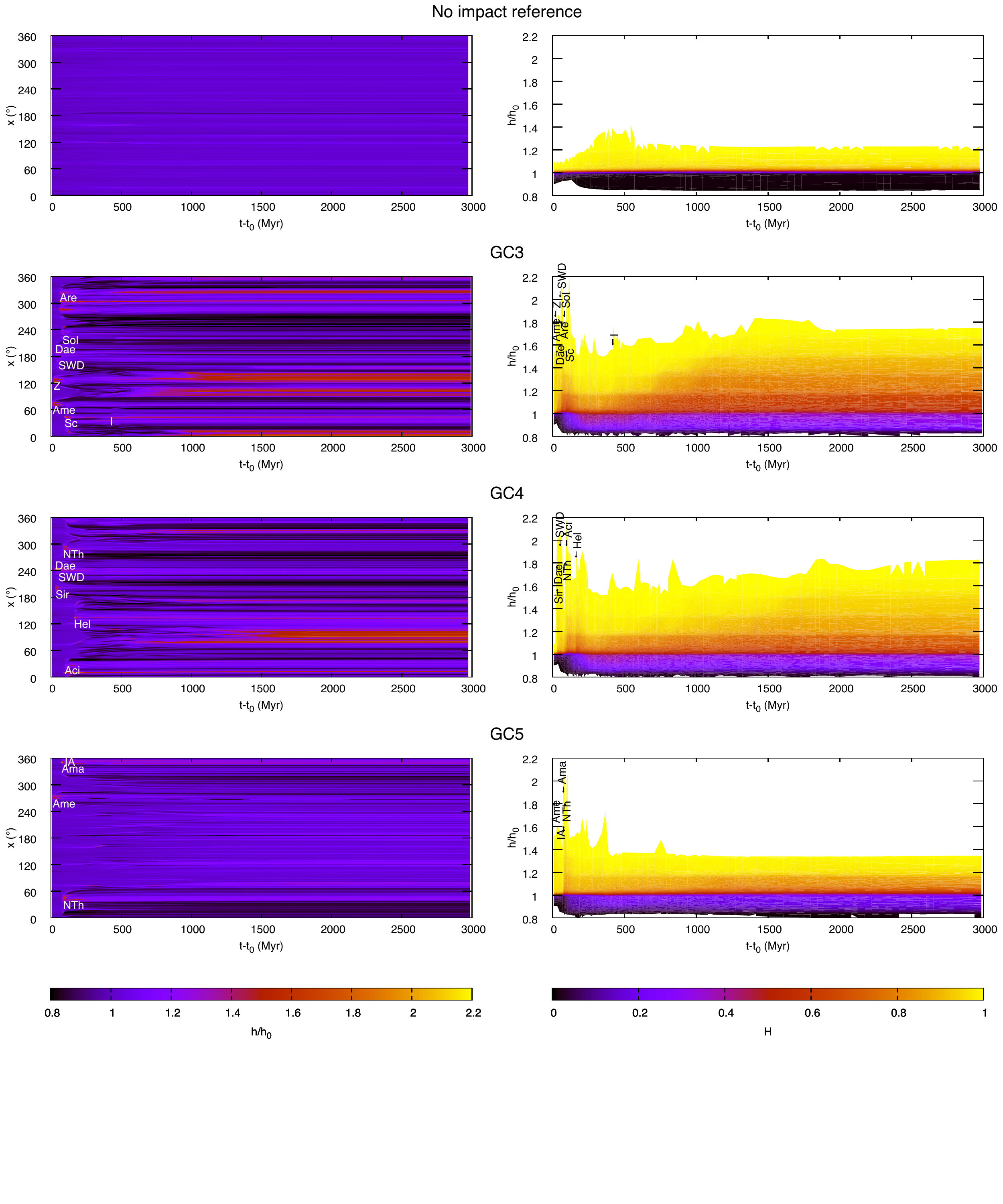}
\caption{Normalized crustal thickness $h(x,t)/h_0$ (left column) and cumulative distribution function $H(h,t)$ of the crustal thickness (right column) of the impact-free reference model and the great-circle models 3, 4, and 5 (cf. Fig.~\ref{fig:vTq-t-gc345} at and after the impacts. The crustal thickness is normalized with the mean crustal thickness $h_0$ immediately before the first impact in Table~\ref{tab:impacts}, i.e., Amenthes at 280\,Myr, which can be calculated from the impact-free reference model. The abscissa gives the time since the first impact; contrary to Fig.~\ref{fig:hcumt-IU}, no normalization has been applied, because the impacts have different sizes. The full set of figures is provided in the Supplementary Material.\label{fig:hcumt-gc345}}
\end{figure}

Given the relatively short lifetime of those temperature-related anomalies, it is of interest to consider how longer-lived anomalies of other characteristics compare between great-circle models. Hence, we look again at the statistics of crustal thickness as expressed by $H(h,t)$ (Fig.~\ref{fig:hcumt-gc345}). As in the two-impact models, we observe that thick crust is formed immediately after the impacts but is partly destroyed by delamination on timescales of a few ten to a few hundred million years. However, the more perturbed dynamics of models with several impacts result in more extensive lithospheric instability in the later evolution, which in turn stirs the mantle up and enables prolonged melt and crust production in some places by transporting relatively fresh or re-enriched material into the melting zone; the refertilization of depleted mantle would be a consequence of the entrainment of delaminated crust \citep[e.g.,][]{Rose:etal14}. As a consequence, there is a second era of (limited and localized) crustal production that begins considerable time after the end of the era of the large impacts and generates some sites of stable thick crust that persist to the present, which can be seen in the $h(x,t)$ plots in the left part of Fig.~\ref{fig:hcumt-gc345}. The resurgent crust production is also visible in some of the $H(h,t)$ plots as a new increase in the maximum thicknesses.\par
For the multi-impact sequences of the great-circle models, it is also worthwhile to take a look at $q_\mathrm{CMB}(t)$, because in most of them we observe substantial lithospheric instability and subsequent transient accumulation of relatively cool material at the CMB. The right panel of Fig.~\ref{fig:Score} shows that indeed in most of these models core entropy production is increased substantially for several hundred million years after the impacts relative to the impact-free reference model, and $q_\mathrm{CMB}$ doubles in this time interval. The only exception, in which the increase is much less pronounced, is model GC5, which had only four impacts and was also the least productive one in terms of post-impact crust formation.

\section{Discussion}
Although the very setup of our models is artificial by design in the two-impact models or due to the restriction in impact count and geometry imposed by the choice of the great circles and the two-dimensional grid, some lessons can be learned with regard to the geological implications of the martian impact record and its uncertainties. There are different compilations of large impact basins in the literature, and they show a substantial variance with regard to the timing, the size, and even the very number of impacts, because many of the old basins that predate Utopia are eroded or otherwise obliterated and therefore difficult to identify and characterize. At the one extreme, the survey by \citet{FrMa13} identified some 30 large basins falling in the time interval between the formation of the dichotomy and the creation of Utopia, some of them of a magnitude comparable to the Utopia event. At the other extreme lies a recent analysis by \citet{BoAn-Ha17} that strongly indicates that such a large flux of giant impactors would almost certainly have hit some part of the dichotomy boundary, but as clear evidence for this seems to be lacking, it concludes that the actual number of such events should be less than 12. We decided to use the list by \citet{Frey08}, because of its intermediate number of basin-forming impacts with potential effects on mantle dynamics and for easier comparison with the study by \citet{JHRobe:etal09}. Moreover, \citet{Werner08} identified partly different basins, assigned them ages that differ by tens or even more than hundred million years from those given in other compilations, and also determined different diameters. On the other hand, the global characteristics we chose to assess the dynamical state tend to converge towards the impact-free reference in the long run; for instance, the final $T_\mathrm{mean}$ in the great-circle models is only about 11--17\,K higher and $q_\mathrm{t}$ less than 1.3\,mW/m$^2$ lower than in the impact-free reference model. This suggests that the uncertainties in the different compilations are not a very strong concern with regard to the overall dynamical and thermal evolution of Mars. In the short-term aftermath of the impacts, however, there are substantial differences, which would leave traces in the geological record that would still be visible today.\par
With regard to the long-term preservation of the impact record, the triggering of lithospheric instability is of particular importance. A crucial factor in this process is the formation of crust that grows thick enough for its base to cross the basalt--eclogite transition and become gravitationally unstable. With our parameters, this transition would generally take place somewhere around 140\,km depth in Mars, which is a bit less than twice the global mean thickness reached after a few hundred million years in the impact-free reference model. That global mean value is rather on the high side of other, independent crustal thickness estimates for Mars and would be lower if a lower initial mantle temperature or a higher melt extraction threshold had been chosen. A lower crustal thickness in our models would imply that lithospheric instability becomes less likely and less extensive or does not occur at all. Likewise, the cooling effect of the delaminated crust would be absent, and models with impacts would always be hotter at all times than the impact-free reference. Thus, post-impact dynamics also depends crucially on pre-impact crust formation and especially on the thickness of the primordial crust, which is poorly known. This highlights once more the importance of obtaining tight constraints on the thickness of the crust, e.g., from the seismic experiment on the InSight mission. Moreover, at the impact sites themselves the actual extent of extraction of melt generated directly by the impacts determines to a large degree the thickness of the newly formed post-impact crust. Our assumption of very efficient extraction is likely an extreme case and defines an upper limit more conducive to crustal delamination. By contrast, \citet{Pado:etal17} assumed no extraction of impact melt and formed their post-impact crust exclusively from melt produced in the aftermath of the impact by the dynamics of the evolving thermal anomaly, which results in a substantially thinner post-impact crust. The true extent of melt extraction in impacts is currently unknown and requires further study.\par
\begin{figure}
\includegraphics[width=\textwidth]{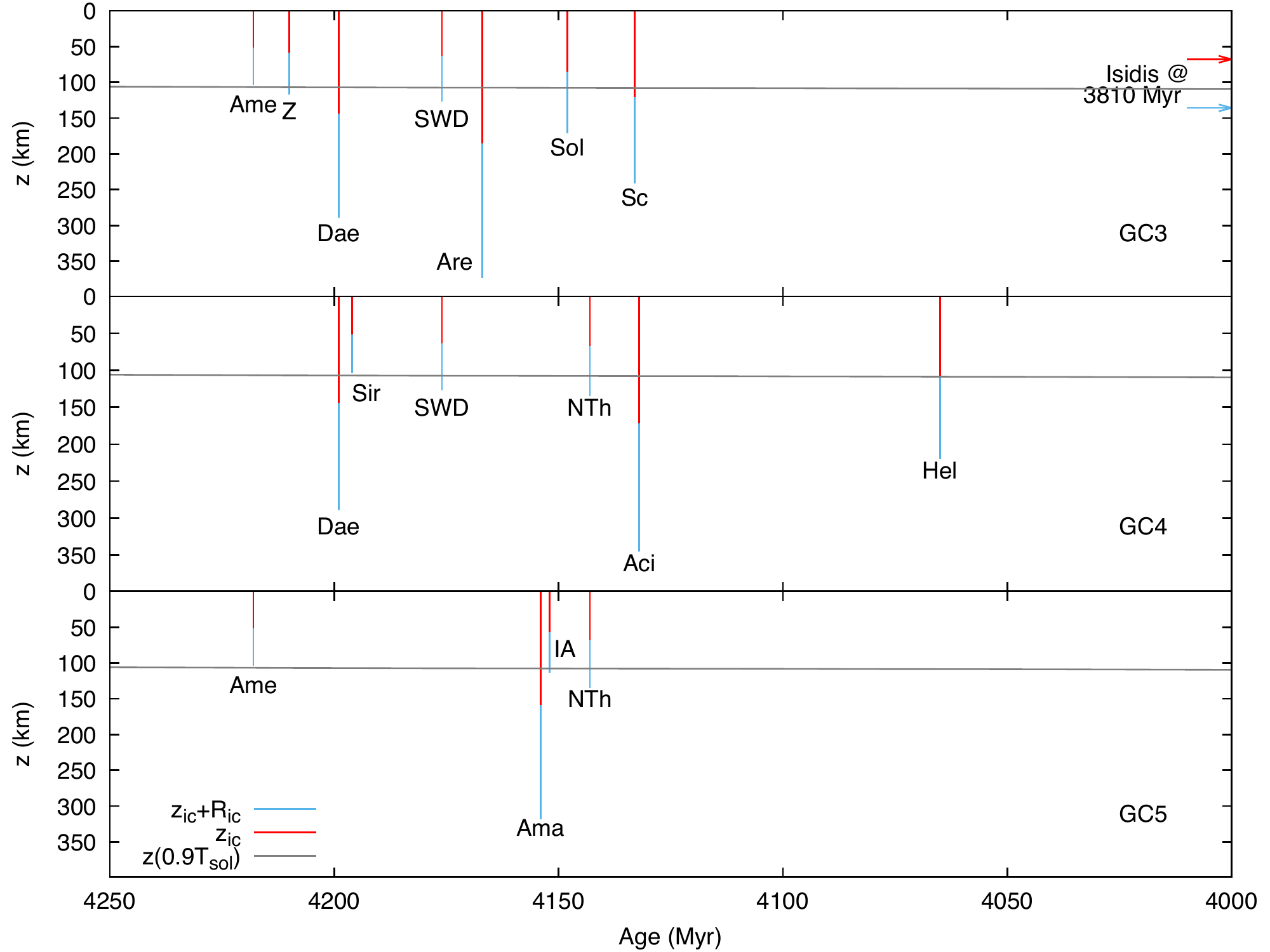}
\caption{Depth range affected by impacts of the great-circle models 3, 4, and 5 in relation to the depth to $0.9T_\mathrm{sol}$, the depth at which 90\% of the solidus temperature of fertile martian peridotite is reached. The red spike indicates the depth $z_\mathrm{ic}$ of the center of the isobaric core, the blue spike the depth to the deepest point inside the isobaric core, $z_\mathrm{ic}+R_\mathrm{ic}$, as a measure of the maximum depth extent of shock-heating. The Isidis event, which occurred much later than all others, is indicated in GC3 with arrows, because it falls outside the plotted time interval. The full set of models can be found in the Supplementary Material.\label{fig:zT90sol}}
\end{figure}

A straightforward criterion for the occurrence of a second melting era, e.g., in terms of a threshold number of impacts or a minimum size does not seem to exist, even though we find that the two-impact models and the great-circle model with the fewest impacts (GC5 with only four events) have no such era whereas the other GC models do. The absence of marked thermal anomalies at the time the second crust-forming era begins suggests that its cause is not primarily thermal; indeed, it is only hundreds of millions of years later that the average mantle temperature of the GC models is again higher than that of the impact-free reference model. The critical ingredient is therefore presumably the availability of easily fusible material such as delaminated crust, which however would first have to be produced from very thick crust by extensive melting. This can happen more easily if the impacts shock-heat an already hot target, which requires them to penetrate more or less the entire lithosphere. Fig.~\ref{fig:zT90sol} shows the depth reach of the impacts in some of the great-circle models as a function of time and in relation to the depth where 90\% of the solidus temperature of fertile martian peridotite is reached; this value serves as a proxy for the shallowest depth at which sublithospheric conditions are reached or at which the mantle is so hot that it will melt without a lot of further energy input. The figure reveals that of the scanty series of impacts in GC5, only one (Amazonis) reaches firmly into the sublithospheric mantle, whereas the other series affect the sublithospheric mantle more strongly and over a longer timespan. On the other hand, the strong effect of the double-Utopia models with two deep-reaching impacts in rather short succession shows that the magnitude alone is not sufficient. Instead, a longer gap between the individual events is more likely to ensure that the convective activity is maintained over a longer period and can stir and mix delaminated crustal material back efficiently. Like all crust-related features, however, this mechanism and hence the very occurrence of the second crust-forming era depends on the efficiency of melt extraction and crust formation during and directly after the impact, as discussed in the previous paragraph.\par
The enhancement of $q_\mathrm{CMB}$ after several large impacts induced by foundered lithospheric material raises the possibility that a core dynamo is reinforced or reactivated by this process, because that cool material causes locally a steep temperature gradient at the CMB that can potentially drive thermal convection in the core. Such an inhomogeneous temperature pattern at the CMB, albeit on a larger scale, had been envisioned by \citet{Stan:etal08} as a possible cause for the dichotomy of the magnetic field of Mars. If the martian crustal dichotomy has been caused by a giant impact \citep[e.g.,][]{Gola:etal11}, however, it is conceivable that similar crust production and eclogite formation would occur on a correspondingly larger scale than in our models, and the effect on the CMB would also be more substantial and global in nature. The models by \citet{Gola:etal11} did not display this effect, but as discussed above, this may be due to the choice of model parameters such as depth to the basalt--eclogite transition or the thickness of the initial crust, which seem to make delamination less likely in their models; as those authors point out, other parameters, e.g., the extraction efficiency or the initial mantle temperature, are also highly uncertain and thus add to the uncertainty in the crustal thickness.\par
Apart from this longer-term, indirect effect of an impact on $q_\mathrm{CMB}$ it is worth mentioning that some workers also observed a slight enhancement of the CMB heat flux over several dozen millions of years directly after the impact as a consequence of the merging of the cores of the impactor and the target planet \citep{MoAr-Ha14}. In principle, modeling this effect requires a more fully dynamical numerical model of core merging than the simple parameterized calculation of heat effects on the mantle used in our models and was therefore omitted; impactors in models based on scaling laws are homogeneous by design. However, the example in figure~16 of \citet{MoAr-Ha14} suggests that even for impactors that are comparable in size or even larger than the largest ones we considered, the effect would not exceed $\sim 5$\,mW/m$^2$ and is thus of a magnitude comparable to that induced by foundering lithospheric material. In impactors of the size and with the physical properties we considered, a core is not expected to exceed a few dozen kilometers in size, and its mass would therefore be at least four orders of magnitude lower than that of the martian core. An order-of-magnitude estimate of the heat from the release of gravitational energy and from the shock that would be transported to the martian core by an intact impactor core suggests that the temperature rise of the core would not exceed a few kelvins even under very favorable conditions. In the more likely scenario of a dispersed impactor core \citep{KeMe16}, the amount of material that eventually ends up in the core will rather be much smaller. Therefore, even in basin-forming impacts core merging is not expected to be significant in terms of the CMB heat flow.\par
\citet{JPWiNi04} used a simple entropy-based criterion for dynamo existence in a non-crystallizing core that considers a cooling rate-dependent entropy term $S_\mathrm{s}$ related to the specific heat of the core, the entropy production from radioactive heating $S_\mathrm{R}$, and the entropy change due to conduction along the core adiabat $S_\mathrm{k}$. Their criterion states that a dynamo can exist if $S_\mathrm{s}+S_\mathrm{R}>S_\mathrm{k}$. The dynamo in our impact-free reference model becomes extinct rather quickly, sooner than commonly thought if the cratering-based timescales for Mars are correct. If some process such as massive crust production and subsequent delamination result in the accumulation of eclogite at the CMB, as is the case as a consequence of impacts in our models, the evolution shown in Fig.~\ref{fig:Score} implies that its influence on the CMB heat flux and hence on core entropy production could extend the lifetime of a dynamo or revive an already defunct one. The resulting timing would also be in better agreement with accepted dynamo durations for Mars, i.e., until $\sim 4.1$\,Ga \citep[e.g.,][]{Lill:etal13b,Verv:etal17}.\par
In the context of discussing global average characteristics of model variables, a few remarks about the characteristics considered here (and in various other studies) will be useful. The issue is the calculation of globally averaged physical properties or variables such as $v_\mathrm{rms}$, $T_\mathrm{mean}$, $q_\mathrm{t,CMB}$, or $h_\mathrm{ave}$ in two-dimensional models and how they relate to the actual three-dimensional geometry of real planets. \citet[App.~B]{RuBr17c} already pointed out that the calculation of variables averaged over the surface such as $h_\mathrm{ave}$ or $q_\mathrm{t}$ results in an overestimate of the signatures of local anomalies such as a crater, because the linear anomaly in a two-dimensional model occupies a larger fraction of the (linear) surface of the model than the two-dimensional crater area does in the two-dimensional surface of a three-dimensional model. The conversion factor for an anomaly spanning an arc $\theta$ from the 2D model to three dimensions is
\begin{equation}
\gamma_{23}=\frac{\pi}{\theta}\left(1-\cos\frac{\theta}{2}\right),
\end{equation}
i.e., it is not constant but depends on size and is greater for anomalies up to about three quarters of the circumference in size. Likewise, anomalous volumes in the mantle affect the average disproportionately in two-dimensional models in comparison to three-dimensional reality. A spherical anomaly with radius $R_\mathrm{a}$ occupies a fraction $f_2=R_\mathrm{a}^2/(R_\mathrm{P}^2-R_\mathrm{c}^2)$ in a 2D model but a fraction $f_3=R_\mathrm{a}^3/(R_\mathrm{P}^3-R_\mathrm{c}^3)$ in a 3D model. The conversion factor for body variables or anomalies is thus
\begin{equation}
\gamma_{23}=\frac{R_\mathrm{P}^2-R_\mathrm{c}^2}{R_\mathrm{P}^3-R_\mathrm{c}^3}R_\mathrm{a}.
\end{equation}
For Mars with the structure assumed here, $\gamma_{23}=2.5163\cdot 10^{-4}R_\mathrm{a}$ (with $R_\mathrm{a}$ in kilometers), i.e., for anomalies of the size of interest here, the amplitudes should be reduced by 0.5 to 1.5 orders of magnitude; correction factors for the other bodies in the inner Solar System are within a factor of about 2. On the other hand, in three dimensions the larger number of impacts would compensate to some extent the downward correction for the geometry effect by $\gamma_{23}$ as far as the sum effect is concerned, because a 3D model with all 20 impacts listed in Table~\ref{tab:impacts} would have between 2.5 and 5 times as many impacts as the 2D models carried out here. This should be kept in mind when comparing absolute values with estimates based on 3D models.

\section{Conclusions}
The thermal and compositional evolution of terrestrial planets is influenced in various ways by large, basin-forming impacts. These impacts are not isolated events but can influence each other via their dynamical effects, if they are close enough to each other in space and time. Very closely spaced impacts occurring shortly after one another can almost appear like a single larger impact, whereas the interaction is less direct and more complex as the distance and/or the time interval between events grows.\par
Under the assumption of complete extraction of the melt generated by the impact, very large impacts can also trigger lithospheric instabilities that modify the convection flow field further, and beyond some threshold these instabilities can become so extensive that they stir the mantle on a global scale and potentially reinforce crustal production, including a late stage of long-lasting but low-volume and localized volcanism. Impact-triggered volcanism also enhances the variability of crustal thickness. The remixing of cool crustal material can lead to an intermediate cooling of the mantle that begins some time after the impacts and lasts several hundreds of millions of years before internal heating of this material partly restores previous temperatures. At this scale, the accumulation of lithospheric material at the CMB may also have a reinforcing effect on core dynamics, in particular on a core dynamo.\par
The most realistic but also most expensive way of modeling mantle dynamics is by a three-dimensional spherical model. Among the consequences of models of a two-dimensional cross-section of a sphere are geometric effects that lead to an overestimate of global averages of certain variables or properties in models which feature strong local anomalies. Two-dimensional models can thus accurately simulate physical processes locally and globally, but in order to derive global average characteristics for a 3D planet such as the mean global heat flow from them, the signatures of strong local processes have to be corrected for the geometry effect. With this caveat, they can still provide useful insights into the evolution of three-dimensional systems.

\section*{Acknowledgments}
Constructive reviews by two anonymous referees helped us to clarify the paper and consider some additional topics of interest. We appreciate helpful comments on size and age estimates for large basins by Stephanie Werner and on some issues related to impact dynamics by Kai Wünnemann. TR was partly supported by DFG (Deutsche Forschungsgemeinschaft) grant Ru 1839/1-1 and supplementary funding from the DFG programme SFB-TRR~170 and DFG grant Ru 1839/2-1. DB was supported by SFB-TRR~170. This is TRR~170 publication no.~56. The numerical calculations were carried out on the computational resource ForHLR~II at the Steinbuch Centre for Computing, Karlsruhe Institute of Technology, funded by the Ministry of Science, Research and the Arts Baden-Württemberg and DFG. Fig.~\ref{fig:gcmap} was made with the Generic Mapping Tools \citep{Wess:etal13}.

\appendix
\setcounter{table}{0}
\setcounter{figure}{0}
\section{Projection of basin positions onto great circles}\label{app:gc}
Real crater locations are given by a longitude and a latitude on a spherical surface, but in two-dimensional models, the impacts are implicitly assumed to occur on a great circle. Here we describe how a suitably chosen subset of impact sites can be fitted with a great circle as closely as possible to determine impact site locations for use in 2D models. We assume that the selected impacts are close enough to the great circle for them to be mapped onto it in a useful manner.\par
For a great circle with a known pole (in the northern hemisphere by convention), this can be achieved by first determining the coordinates of the circle's northernmost point:
\begin{equation}
\lambda_\mathrm{N}=\lambda_\mathrm{P}-\pi,\qquad
\varphi_\mathrm{N}=\frac{\pi}{2}-\varphi_\mathrm{P},
\end{equation}
where $\lambda$ and $\varphi$ are longitude and latitude, and the subscripts P and N indicate the pole of the great circle and its northernmost point, respectively. The equations of the great circle and its derivative with respect to $\lambda$ are then
\begin{subequations}
\begin{align}
\varphi&=\arctan[\tan\varphi_\mathrm{N} \cos(\lambda-\lambda_\mathrm{N})]\\
\drv{\varphi}{\lambda}&=\frac{-\tan\varphi_\mathrm{N} \sin(\lambda-\lambda_\mathrm{N})}{1+[\tan\varphi_\mathrm{N} \cos(\lambda-\lambda_\mathrm{N})]^2}=
\frac{-\tan\varphi_\mathrm{N} \sin(\lambda-\lambda_\mathrm{N})}{1+\tan^2\varphi}.\label{eq:dphidl}
\end{align}
\end{subequations}
The distance $\delta$ from the actual impact location $(\lambda_\mathrm{c},\varphi_\mathrm{c})$ to a point $(\lambda,\varphi)$ on the great circle is given by the orthodrome arc
\begin{equation}
\delta=\arccos[\sin\varphi_\mathrm{c}\sin\varphi+\cos\varphi_\mathrm{c}\cos\varphi\cos(\lambda-\lambda_\mathrm{c})],
\end{equation}
and should be minimal for the projection point on the great circle, i.e., its derivative with respect to the longitude $\lambda$ should be zero:
\begin{align}
\drv{\delta}{\lambda}&=-\frac{\sin\varphi_\mathrm{c}\cos\varphi\drv{\varphi}{\lambda}-\cos\varphi_\mathrm{c}\left[\sin\varphi\drv{\varphi}{\lambda}\cos(\lambda-\lambda_\mathrm{c})+\cos\varphi\sin(\lambda-\lambda_\mathrm{c})\right]}{\sqrt{1-[\sin\varphi_\mathrm{c}\sin\varphi+\cos\varphi_\mathrm{c}\cos\varphi\cos(\lambda-\lambda_\mathrm{c})]^2}}\nonumber\\
&=-\frac{\sin\varphi_\mathrm{c}\cos\varphi\drv{\varphi}{\lambda}-\cos\varphi_\mathrm{c}\left[\sin\varphi\drv{\varphi}{\lambda}\cos(\lambda-\lambda_\mathrm{c})+\cos\varphi\sin(\lambda-\lambda_\mathrm{c})\right]}{\sin\delta}=0,
\end{align}
where $\drv{\varphi}{\lambda}$ is taken from Eq.~\ref{eq:dphidl}. The two existing solutions correspond to the maximum and the minimum distance. The desired minimum solution is found numerically, whereby the maximum distance can easily be excluded by applying the root-finding algorithm in a sufficiently narrow interval around $\lambda_\mathrm{c}$, assuming that the great circle is a good enough fit. The coordinate for use in the model can then be given as the arc distance of the projection point $(\lambda'_\mathrm{c},\varphi'_\mathrm{c})$ from the northernmost point of the great circle as
\begin{equation}
s=\arccos[\sin\varphi_\mathrm{N}\sin\varphi'_\mathrm{c}+\cos\varphi_\mathrm{N}\cos\varphi'_\mathrm{c}\cos(\lambda'_\mathrm{c}-\lambda_\mathrm{N})].
\end{equation}
The placement of the impacts on a great circle in our GC model series uses this relation. Hence, a given impact will have different positions in different models.

\section{Poiseuille flow with temperature-dependent rheology}\label{app:pois}
In order to construct a model simple enough for an analytical solution that captures some features of the upwelling in a hot plume generated by the thermal anomaly of an impact, we assume: 1. that the flow is steady and occurs in a vertical cylindrical pipe with rigid walls at $\pm R$ as in the classical pipe-flow (Poiseuille flow) problem; 2. that the temperature anomaly $\Delta T$ is only a function of distance $r$ from the central axis of the pipe; 3. that the decisive influence on viscosity is the temperature and that it can be described by a Frank-Kamenetskii law, i.e., $\eta(T(r))=\eta_0 \exp(-b\Delta T(r))$ with constant parameters $\eta_0$ and $b$; 4. that the timescale of the flow process is much shorter than the thermal diffusion timescale, such that we can assume a time-independent $\Delta T(r)$ that is zero for all $r\geq R$ and $\Delta T_\mathrm{max}$ at $r=0$. For our purposes, $\Delta T(r)=\Delta T_\mathrm{max}[1-(r/R)^2]$ is a useful choice that fulfills these criteria.\par
We start with the same assumptions as in the derivation for the classical Poiseuille flow problem as given by \citet{TuSc82}, i.e., by stating that the force resulting from the pressure difference $\Delta p$ that drives the flow and the shear force $\tau$ on the walls of the cylindrical control volume of length $l$ and radius $r$ cancel out:
\begin{equation}
\pi r^2 \Delta p=-2\pi rl\tau \Leftrightarrow
\tau=\frac{r}{2}\frac{\mathrm{d}p}{\mathrm{d}z}.\label{eq:fbal}
\end{equation}
In our setup, $\tau$ is directly related to the flow velocity $v$ by the flow law, i.e., $\tau=\eta \mathrm{d}v/\mathrm{d}r$. The driving pressure results from the thermal buoyancy of the anomaly, which causes a contrast in the vertical lithostatic pressure gradient of $\mathrm{d}p/\mathrm{d}z=g\Delta\varrho(r)$. The density variation is controlled by the thermal expansion due to the temperature anomaly, i.e., $\Delta\varrho=-\varrho_0 \alpha \Delta T(r)$; $\varrho_0$ and $\alpha$ are the background density and thermal expansivity, respectively, and are constant. Setting Eq.~\ref{eq:fbal} equal to the flow law and inserting the relations for the viscosity, the density and temperature anomaly, and the pressure gradient, we arrive at a differential equation that can be solved by separation of variables:
\begin{gather}
\eta_0 \exp\left\lbrace -b\Delta T_\mathrm{max}\left[1-\left(\frac{r}{R}\right)^2\right]\right\rbrace \frac{\mathrm{d}v}{\mathrm{d}r}=\frac{g\varrho_0 \alpha \Delta T_\mathrm{max} r}{2}\left[1-\left(\frac{r}{R}\right)^2\right]\\
\Rightarrow v(r)=-\frac{g\varrho_0 \alpha R^2}{4\eta_0 b^2 \Delta T_\mathrm{max}}\left\langle \left\lbrace
1-b\Delta T_\mathrm{max}\left[1-\left(\frac{r}{R}\right)^2\right]\right\rbrace\exp\left\lbrace b\Delta T_\mathrm{max} \left[1-\left(\frac{r}{R}\right)^2\right]\right\rbrace -1\right\rangle.
\end{gather}
This flow profile resembles a Gaussian bell curve and is more focused at the axis for a given $R$ than the classical Poiseuille flow profile, which is parabolic. The focusing increases with the magnitude of the temperature anomaly and with increasing $b$.

\section{Simplified anomaly volumes}\label{app:anoVA}
In order to estimate the dependence of the impact-generated thermal anomalies on their spatial and temporal separation, we need the dependence of the anomaly volumes on the separation of the two impacts, $\Delta x$. As we are only interested in a general order-of-magnitude estimate, we approximate the anomalies by two overlapping spheres of radius $R$ whose centers are $\Delta x$ apart from each other, i.e., the anomalies are centered at $x=R$ and $x=R+\Delta x$ and intersect at $x=R+\Delta x/2$; $R$ can be any reasonable length scale of the anomaly, likely a small multiple of the isobaric core radius. Our anomaly outline is thus defined by the outline
\begin{equation}
a(x)=
\begin{cases}
\sqrt{R^2-(x-R)^2}=\sqrt{2xR-x^2},&0\leq x\leq R+\Delta x/2\\
\sqrt{R^2-(x-R-\Delta x)^2},&R+\Delta x/2<x\leq 2R+\Delta x/2
\end{cases}
\end{equation}
and zero elsewhere. With respect to the comparison of two- and three-dimensional models, we derive the solutions for both two- and three-dimensional anomalies. As both anomalies are assumed to be of equal size, all integrals are only calculated on the interval $[0; R+\Delta x/2]$, and the result is doubled. We only need to consider separations up to $2R$, because beyond that distance, there are simply two separate spherical anomalies.\par
The ``volume'' of the anomaly in two-dimensional models such as our numerical models is the area between $a(x)$ and $-a(x)$:
\begin{align}
V&=4\int\limits_0^{R+\frac{\Delta x}{2}} \sqrt{2xR-x^2}\,\mathrm{d}x=R^2 \left[\frac{\Delta x}{R}\sqrt{1-\frac{\Delta x^2}{4R^2}}+2\arcsin\left(\frac{\Delta x}{2R}\right)+\pi\right]\nonumber\\
&=A_\mathrm{circ} \left[1+\frac{\Delta x}{\pi R}\sqrt{1-\frac{\Delta x^2}{4R^2}}+\frac{2}{\pi}\arcsin\left(\frac{\Delta x}{2R}\right)\right],\label{eq:V-2D}
\end{align}
where $A_\mathrm{circ}$ is the area of a circle with radius $R$.\par
In a three-dimensional real planet, the idealized anomaly is given by rotating $a(x)$ around the $x$ axis. The rotational body has the volume
\begin{align}
V&=2\int\limits_0^{R+\frac{\Delta x}{2}} \pi a^2(x)\,\mathrm{d}x=
2\pi\int\limits_0^{R+\frac{\Delta x}{2}} (2xR-x^2)\,\mathrm{d}x\nonumber\\
&=\pi\left(\frac{4}{3}R^3+R^2\Delta x-\frac{1}{12}\Delta x^3\right)=
V_\mathrm{sph}\left[1+\frac{3}{4}\frac{\Delta x}{R}-\frac{1}{16}\left(\frac{\Delta x}{R}\right)^3\right],\label{eq:V-3D}
\end{align}
where $V_\mathrm{sph}$ is the volume of a sphere with radius $R$. Both functions are strictly monotonic on the interval of interest, $0\leq x/R\leq 2$. In their final forms, they are cast as products of the volume of a simple sphere (in two or three dimensions) and a form factor. The form factor functions for volumes differ by at most 5\% between the two- and three-dimensional cases, which suggests that effects that depend on volume follow comparable trends in their dependence on impact spacing in two and three dimensions.\par
In this discussion, we have also assumed that the spheres are fully immersed in the target planet, i.e., they are not truncated by the surface. In the models, we have set the depth of the isobaric core according to the estimate
\begin{equation}
z_\mathrm{ic}=0.1524 D_\mathrm{imp} v_\mathrm{imp}^{0.361}\left(\frac{\varrho_\mathrm{imp}}{\varrho}\right)^{0.5}
\end{equation}
\citep[eq.~6]{RuBr18b}, which is an extension of an empirical relation by \citet{Pier:etal97}. Comparison of the isobaric core diameters and depths in Table~\ref{tab:impacts} shows that this simplification is well justified for the impacts considered here if $R$ is taken to be of the order of $R_\mathrm{ic}$.

\section{Effect of crater overlap on heat flow}\label{app:qcrat}
We consider only complex craters and neglect the curvature of the target body for simplicity. Taking the diameter of the final crater as the measure of the area in which the heat flow is modified, two (identical) impacts will be independent if the distance of their centers is greater than $D_\mathrm{f}$. For a given impact, our measure of the isobaric core is given by
\begin{equation}
D_\mathrm{ic}=2r_\mathrm{infl}=D_\mathrm{imp}\left(\frac{n-1}{n+1}b\right)^\frac{1}{n},\label{eq:Dic}
\end{equation}
where $b$ and $n$ are empirical parameters that depend on $v_\mathrm{imp}$ \citep{Ruedas17a}. Solving for $D_\mathrm{imp}$, equating with Eq.~\ref{eq:DimpDf}, and rearranging gives
\begin{equation}
D_\mathrm{f}=1.3836\left(\frac{\varrho_\mathrm{imp}}{\varrho}\right)^{0.377} \frac{v_\mathrm{imp}^{0.4972}}{g^{0.2486}D_\mathrm{sc}^{0.13}}\left(\frac{n+1}{b(n-1)}\right)^\frac{0.8814}{n} D_\mathrm{ic}^{0.8814},
\end{equation}
i.e., $D_\mathrm{f}\sim D_\mathrm{ic}^{0.8814}$; for our setup $b=2.463$ and $n=1.424$, and hence $D_\mathrm{f}\approx 17.9 D_\mathrm{ic}^{0.8814}$. Thus, all of our double-impact models except the Utopia-sized ones with $\Delta x'=10$ are expected to lie in the overlap region. It should be kept in mind, however, that $D_\mathrm{f}\gg D_\mathrm{ic}$ does not imply that the entire thermal anomaly is shielded by the modified crust; the shock-heated region is several times larger than the isobaric core, in which the highest shock pressures are reached.\par
Neglecting curvature, we can use Eq.~\ref{eq:V-2D} with $R$ being replaced by $D_\mathrm{f}/2$ to estimate the effective total impact-modified area for overlapping impacts on a sphere:
\begin{equation}
A_q=\frac{D_\mathrm{f}^2}{4} \left[\pi+\frac{2\Delta x}{D_\mathrm{f}}\sqrt{1-\frac{\Delta x^2}{D_\mathrm{f}^2}}+2\arcsin\left(\frac{\Delta x}{D_\mathrm{f}}\right)\right].
\end{equation}
For our two-dimensional models, the modified ``surface'' is an arc of length $x\geq D_\mathrm{f}$, and so the effective modified ``area'' increases linearly with $\Delta x/D_\mathrm{f}$ up to $2D_\mathrm{f}$.


\end{document}